\documentclass[a4paper,12pt]{article}
\usepackage{jheppub} 
\usepackage{braket}
\usepackage{subfigure}
\usepackage{simplewick}
\usepackage{mathrsfs, extarrows, amssymb}
\newcommand\td{\text{d}}
\newcommand{\p}{\partial}

\def\>{\rangle} \def\<{\langle}

\title{\boldmath QFT in Klein space} 
\author[a,b,c,d]{Bin Chen}
\author[b]{Zezhou Hu}
\author[b]{Xin-Cheng Mao}

\affiliation[a]{Institute of Fundamental Physics and Quantum Technology, \\\&  School of Physical Science and Technology, \\ Ningbo University, Ningbo, Zhejiang 315211, China}
\affiliation[b]{School of Physics, Peking University, \\No.5 Yiheyuan Rd, Beijing 100871, P.~R.~China}
\affiliation[c]{Center for High Energy Physics, Peking University, \\No.5 Yiheyuan Rd, Beijing 100871, P.~R.~China}
\affiliation[d]{Peng Huanwu Center for Fundamental Theory, \\ Hefei, Anhui 230026, P. R. China}
\emailAdd{chenbin1@nbu.edu.cn, z.z.hu@pku.edu.cn, 
maoxc1120@stu.pku.edu.cn}

\abstract{In this paper, we investigate the quantum field theory in Klein space that has two time directions. To study the canonical quantization, we select the ``length of time" $q$ as the evolution direction of the system. In our novel construction, some additional modes beyond the plane wave modes are crucial in the canonical quantization and the later derivation of the LSZ reduction formula. We also derive the free two-point function by using Wick contraction in the canonical quantization formalism. Moreover, we introduce the path-integral formalism in which we can redefine the vacuum states and rederive the correlation functions. We show that all the results in the Klein space derived in our novel approach match those obtained via analytical continuation from the Minkowski spacetime.}

\begin{document}
\maketitle
\flushbottom
%\newpage

\section{Introduction}

The holographic principle \cite{tHooft:1993dmi, Susskind:1994vu} is the cornerstone of quantum gravity. It suggests a duality between quantum gravity in bulk and quantum field theory at a lower-dimensional boundary. A prototypical example of this duality is the celebrated AdS/CFT correspondence \cite{Maldacena:1997re, Gubser:1998bc, Witten:1998qj}.  To deepen our understanding of the holographic principle, significant efforts have been devoted to extending the AdS/CFT correspondence to flat spacetime holography, as explored in early works \cite{Susskind:1998vk, Polchinski:1999ry,deBoer:2003vf, Arcioni:2003xx, Arcioni:2003td, Solodukhin:2004gs, Barnich:2006av, Guica:2008mu, Barnich:2009se, Barnich:2010eb, Bagchi:2010zz, Bagchi:2012xr}. In recent years,  flat holography in Minkowski spacetime has been widely studied in the framework of celestial holography \cite{Pasterski:2016qvg, Pasterski:2017kqt, Pasterski:2017ylz, Raclariu:2021zjz, Pasterski:2021rjz, Pasterski:2021raf, Strominger:2017zoo} and the Carrollian perspective \cite{Donnay:2022aba, Donnay:2022wvx, Chen:2023naw, Chen:2023pqf, Bagchi:2023cen, Bagchi:2022emh, Bagchi:2024unl}.

Holography in flat spaces $\mathbb{R}^{n,m}$  with varying $n, m$ is closely related to the AdS/CFT correspondence. On one hand, Carrollian holography in Minkowski space emerges from the flat limit of AdS/CFT \cite{Bagchi:2012cy, Bagchi:2022nvj, Banerjee:2022ime, Bagchi:2023fbj, Alday:2024yyj}. Moreover, techniques developed in AdS/CFT can be applied  directly to celestial holography in Minkowski \cite{Ball:2019atb, Casali:2022fro, Iacobacci:2022yjo, Melton:2023dee, Bu:2023cef}, %Bu:2023vjt, Bu:2024cql, 
since the Minkowski space can be foliated by Euclidean AdS slices inside the lightcone of the origin. On the other hand, AdS spaces AdS$_q$ themselves can be viewed as hyperbolic slices embedded in higher-dimensional flat spaces $\mathbb{R}^{2,q}$ with two time directions. In particular, AdS$_3$ slices share conformal boundaries with such embeddings, suggesting that flat holography in two-time space $\mathbb{R}^{2,2}$ may shed new light on AdS$_3$/CFT$_2$. For concreteness, this work focuses on the simplest multi-time prototype: Klein space $\mathbb{R}^{2,2}$ (or $\mathbb K^{2,2}$).

%since the AdS spaces are nothing but a series of hyperbolic slices embedded in flat spaces with two time directions\footnote{Neither of $q$ nor $p$ for multi-time space can be 1. This is because manifolds $\mathbb{R}^{p,q}$ with one time or one space direction are mathematically equivalent.} that share the same boundary with the AdS$_3$ slices, researches on flat holography within the two-time flat space can also be beneficial for studying the AdS/CFT correspondence. Without losing universality, this paper will focus on the simplest example of multi-time flat space, known as the Klein space $\mathbb{R}^{2, 2}$ . %The metric of Klein space is given by
%\begin{equation}
    %\td s^2 = -(\td x^0)^2-(\td x^1)^2 + (\td x^2)^2 + (\td x^3)^2\,.
%\end{equation}

The scattering amplitudes in  Klein space present some intriguing features that are trivial in Minkowski and Euclidean spaces. %Although the Klein space can be related to Minkowski through Wick rotations, the physics in Klein space may still have some differences from the discussions in Minkowski space. 
For instance, the on-shell three-point amplitudes of massless particles do not exist in Euclidean space due to the nonexistence of on-shell states, and they universally vanish in Minkowski spacetime for kinematic reasons, but they are non-trivial in Klein space. Therefore, based on the demonstrations in \cite{Britto:2005fq, Cachazo:2004kj, Arkani-Hamed:2008bsc, Benincasa:2007xk, Arkani-Hamed:2012zlh}, the massless scattering amplitudes in Klein space can be expressed in terms of three-point amplitudes, instead of relying on four-point amplitudes as in Minkowski or Euclidean space. This results in a simpler structure of scattering amplitudes in Klein space. On the other hand, modern approaches for dealing with scattering amplitudes through complexification can be much simpler in Klein space than in Minkowski or Euclidean space, see, e.g. \cite{Penrose:1967wn, Penrose:1968me, Penrose:1985bww, Parke:1986gb, Dunajski:2001ea, Witten:2003nn, Arkani-Hamed:2009hub, Monteiro:2020plf}. 

Klein space has a notable feature that there is only one single asymptotic conformal boundary, in contrast to Minkowski spacetime's two boundaries. This unique property makes Klein space particularly suitable and convenient for investigations of flat holography, especially in the context of recent developments in celestial holography \cite{Atanasov:2021oyu, Melton:2023hiq, Melton:2023bjw, Melton:2024jyq, Melton:2024pre, Duary:2024cqb, Bhattacharjee:2021mdc}. Based on all these aforementioned developments, the Klein space serves as a valuable tool, providing new insights into the intricate features of the theories in the Lorentz signature, including self-dual quantum gravity \cite{Penrose:1976jq, KO198151, flathspaces, Eguchi:1978xp}, black holes \cite{Crawley:2021auj}, the Unruh effect \cite{Santos:2023pwg}, etc.
%Furthermore, Klein space also plays an important role in many other contexts, such as self-dual quantum gravity \cite{Penrose:1976jq, KO198151, flathspaces, Eguchi:1978xp}, black holes \cite{Crawley:2021auj}, etc. %the FLRW cosmology
%string theory with extended $\mathcal N = 2$ worldsheet supersymmetry \cite{Ooguri:1990ww, Ooguri:1991fp, Ooguri:1991ie}, F-theory \cite{Vafa:1996xn}

%Beyond studying holography, the physics in the bulk of Klein space is also important. 
%In fact, the above two features appear in the more general spacetime $\mathbf{R}^{p,q}$ with $p,q\ge2$. In this paper, we will focus on the simplest example when $p=q=2$, that is the Klein space.

%\hzz{More motivation on why we study Klein space is needed here. More introduction on Klein space is needed here. We also need some graphs to demonstrate the physical picture.}

%Based on all these previous developments on Klein space, it is meaningful to investigate the scattering theories for quantum field theory (QFT) in it. 

If the two time directions in the Klein space are treated as independent evolution parameters, one inevitably introduces two conjugate momenta, giving rise to unphysical degrees of freedom and additional ghost modes even for the free scalar field. This seems to spoil the causality and unitarity. Fortunately, the problem has been solved after introducing an additional local gauge symmetry acting on phase space to eliminate the unphysical degrees of freedom originating from the extra time dimension. This kind of discussion is referred to as 2T-physics. For more details, see \cite{Bars:1998ph, Bars:2000qm} and the references therein, see also more recent progress in \cite{Bars:2010xi, Bars:2020mad}. By making the additional gauge choice and solving the associated constraint, one obtains a 1T physics with lower spacetime dimensions. As a result, the causality and unitarity are preserved. In such a setting, the associated 1T-physics does not have to be Lorentz invariant; it can be non-relativistic.

In contrast, in this paper, we have a totally different physical setup. We pick out only one time direction as the evolution parameter, leading to a single conjugate momentum and the absence of additional ghosts. Unlike 2T-physics, in our approach, the spacetime symmetry plays a central role, since it can be viewed as a reformulation of relativistic QFT, turning the original ISO(1,3) spacetime symmetry to ISO(2,2). It is therefore reasonable to expect that the present formulation can be related to conventional QFT in Minkowski spacetime through analytic continuation.

A standard QFT is formulated in Minkowski space to calculate the scattering amplitude of ingoing and outgoing particles from the past and the future boundaries, based on the path-integral quantization and canonical quantization. The analytic continuation from Lorentz signature to Kleinian signature is natural in the path-integral language \cite{Heckman:2022peq}, similar to previous investigations of analytic continuation to Euclidean signature \cite{Schwinger:1958mma}.
%As a simple and illuminating example, the quantum field theory of scalar fields in Klein space is worth exploring.
However, since canonical quantization %(reviewed in appendix \ref{ap:CanonicalQ}) 
relies on a 3+1 decomposition of spacetime, it is subtle to do canonical quantization in Klein space, as it has two timelike directions. The most peculiar aspect is that when one of the time directions is selected for decomposition, the symmetry between the two time directions gets broken.
%it becomes more special than the other. This contradicts the expectation that both time directions should be symmetric, with neither being more special than the other.

In this paper, to keep the symmetry of the two time directions, we will use the coordinate $q$, the ``length" of the time, as ``time" (the evolution parameter) to investigate the canonical quantization of the scalar field and its corresponding scattering theory. Because of the existence of only one single conformal boundary, the scattering process in Klein space cannot be described by an $S$-matrix in a normal way, instead, it takes the form like a vector, referred to as $S$-vector. The physical meanings of $S$-vector and furthermore the ``time-ordering'' in correlation functions has not been discussed carefully in the literature.\footnote{In \cite{Melton:2024pre}, the authors constructed the $S$-vector as a Poincar\'e
invariant vacuum state $|\mathcal{C}\>$ in the Hilbert space built on $\mathcal{J}$, where the Hilbert space $\mathcal{H}_\mathcal{J}=\mathcal{H}_{\mathcal{J}_+}\otimes\mathcal{H}_{\mathcal{J}_-}$. And the Hilbert spaces $\mathcal{H}_{\mathcal{J}_\pm}$ are built on AdS$_3/\mathbb{Z}$ slices, which is quite different from our starting point. More importantly, they did not explicitly show that their constructions lead to the results obtained from the analytical continuation from Minkowski spacetime.}  We would like to address these issues in this work.

%In the literature, 2T physics \cite{Bars:1998ph, Bars:2000qm, Bars:2010xi, Bars:2020mad} has been studied broadly. The 2T physics refers to the physics with two temporal directions, based on additional gauge symmetry acting on phase space. One can make a gauge choice and solve the associated constraint to obtain a 1T physics with lower spacetime dimensions. In such a setting, the associated 1T-physics does not have to be Lorentz invariant; it can be non-relativistic. However, spacetime symmetry plays an important role in our paper. Thus, to avoid any confusion, we state that our construction is not a 2T physics but a reformulation of QFT with Lorentz spacetime symmetry, although the Klein space also has two temporal directions.

The remaining parts of this paper are organized as follows. In section \ref{sec:RevKlein}, we review briefly the basic knowledge about the Klein space and especially introduce the coordinate system we will use when performing quantization. Then in section \ref{sec:CanonicalQ}, we perform canonical quantization by expanding the scalar field in terms of Bessel functions and Neumann functions, and define the Neumann vacuum state and the Hankel vacuum state. Within the ``in-out" formalism, we explicitly calculate the free two-point function. We also derive the LSZ reduction formula for a general interacting field theory. 
In section \ref{sec:PathInt}, we interpret the states and $q$-ordered correlation functions in the framework of the path-integral quantization and rederive the free two-point function. Finally, we draw conclusions and discuss the implications of our work in section \ref{sec:ConDiss}. In appendix \ref{ap:analyCon}, we show that the results obtained in our novel approach exactly match the results by direct analytical continuation from Minkowski spacetime. In appendix \ref{ap:propagator}, we check the self-consistency of the propagator.  In appendix \ref{ap:PertExpand}, we show how to do perturbative calculation in Klein space. In addition, we demonstrate the causal structure of the correlation functions within a very simple example, the two-point function of conformal field theory in Klein space, in the appendix \ref{ap:CausalStr}. Some discussions on the QFT with more time directions can be found in our work \cite{Chen:2025eeh}.

\section{Basics of Klein space} \label{sec:RevKlein}

%3+1 decomposition in Klein space is necessary in canonical quantization. In this paper, we will adopt the formalism that the Klein space is foliated by the "length of time" $q$.
In this section, we  briefly introduce the fundamental aspects of Klein space $\mathbb K^{2, 2}$, including its geometric structure,  conformal boundaries, and dual space (known as the momentum space). Let us start from its Penrose diagram.

In Cartesian coordinates, the flat metric of $\mathbb K^{2, 2}$  takes the form:
\begin{equation}
    \td s^2 = - (\td x^0)^2 - (\td x^1)^2 + (\td x^2)^2 + (\td x^3)^2.
\end{equation}
The Klein space can be decomposed into two orthogonal 2D planes, each of which can be parameterized using polar coordinates
\begin{equation}
    (x^0, x^1) = q (\cos \psi, \sin \psi), \qquad (x^2, x^3) = r (\cos \varphi, \sin \varphi),
\end{equation}
such that the metric can be rewritten as
\begin{equation}
    \td s^2 = - \td q^2 - q^2 \td \psi^2 + \td r^2 + r^2 \td \varphi^2.
\end{equation}
Then, with the following coordinate transformations
\begin{equation}
    Q = \arctan (q+r) + \arctan (q-r), \quad R = \arctan (q+r) - \arctan (q-r),
\end{equation}
 the metric can be rewritten as 
\begin{equation}
    \td s^2 = \Omega^{-2} (- \td Q^2 + \td R^2 - \sin^2 Q \td \psi^2 + \sin^2 R \td \varphi^2), %\frac{1}{4 \cos^2 (\frac{R + Q}{2}) \cos^2 (\frac{R - Q}{2})},
\end{equation}
where the conformal factor is defined as
\begin{equation}
    \Omega = 2 \left| \cos \left( \frac{R + Q}{2} \right) \cos \left( \frac{R - Q}{2} \right) \right|.
\end{equation}
The infinite coordinate ranges $q, r \in [0, \infty)$ are conformally mapped to finite ranges $Q, R \in [0, \pi)$. The conformal boundary emerges at the singularity of $\Omega^{-1}$, with null infinity $\mathscr{I}$ located at $Q + R = \pi$. The endpoints $(Q, R) = (0, \pi)$ and $(Q, R) = (\pi, 0)$ correspond to the spacelike infinity $i_0$ and timelike infinity $i'$, respectively.
\begin{figure}[htbp]
    \centering
    \includegraphics[width=0.5\linewidth]{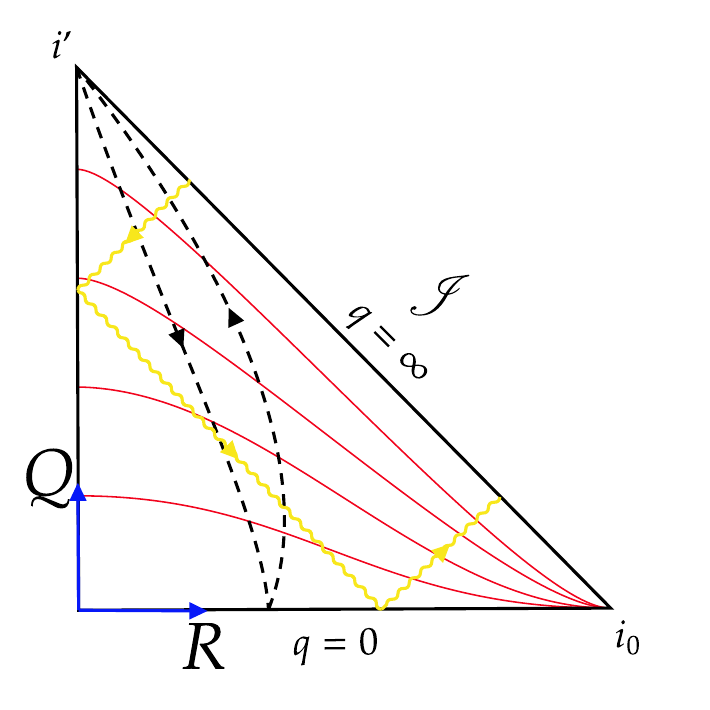}
    \caption{The Penrose diagram of Klein space.}
    \label{fig:KleinPen}
\end{figure}
The Penrose diagram can thus be depicted in the $(Q, R)$ coordinates, as shown in Fig. \ref{fig:KleinPen}. In this diagram, the black dashed line and the yellow wavy line (45°) denote the worldlines of massive and massless particles, respectively. Unlike the Minkowski space $\mathbb R^{1,3}$, which has two  null boundaries, the Klein space possesses only a single null infinity. This structural difference necessitates replacing the conventional S-matrix with an S-vector to describe scattering processes, following the discussion in \cite{Witten:2001kn}.

The Klein space admits a natural foliation by $q = $ constant hypersurfaces, represented by red curves in Fig \ref{fig:KleinPen}, where the $q  = 0$ limit approaches the hypersurface $Q = 0$ and $q  \rightarrow \infty$ limit converges to null infinity $\mathscr{I}$. For our subsequent analysis, we employ polar coordinates for the temporal dimensions only while maintaining Cartesian coordinates for the spatial dimensions. This hybrid coordinate system proves particularly advantageous when examining  the $q = $ constant hypersurface. Under this choice, the metric takes the form
\begin{equation} \label{eq:polar}
    \td s^2 = - \td q^2 - q^2 \td \psi^2 + (\td x^2)^2 + (\td x^3)^2.
\end{equation}

The momentum space, as a dual space of Klein space, inherits the same geometric structure as $\mathbb K^{2,2}$. Consequently, we can parameterize the 4-momentum using analogous coordinates:
\begin{equation}
    p^\mu=(p^0,p^1,p^2,p^3)=(\omega \cos\theta,\omega \sin\theta,\vec{p})\,,\quad \vec{p}=(p^2,p^3)\,,\quad \omega>0.
\end{equation}
Then the $SO (2, 2)$-invariant momentum measure can be defined in the similar manner as that in Minkowski space
\begin{equation}
    \int \td^4 p \delta (p^2 + m^2) %& = \int \sqrt{g} \td \omega \td \theta \td^2 \vec p \delta (- \omega^2 + \vec p^2 + m^2) \\
    = \int \omega \td \omega \td \theta \td^2 \vec p \delta (- \omega^2 + \omega_{\vec{p}}^2) = \frac12 \int^{2 \pi}_0 \td \theta \int \td^2 \vec p \bigg|_{\omega = \omega_{\vec{p}}}
\end{equation}
where
\begin{equation}
    \omega = \sqrt{(p^0)^2 + (p^1)^2}, \qquad \omega_{\vec{p}} = \sqrt{|\vec p|^2 + m ^2}, \qquad |\vec p| = \sqrt{(p^2)^2 + (p^3)^2}.
\end{equation}
The $\vec p$ denotes the spacial momentum, and the $\omega_{\vec p}$ is  the on-shell energy.

\section{Canonical quantization for real scalar in Klein space} \label{sec:CanonicalQ}

The canonical quantization of field theories in Minkowski spacetime is fundamentally based on a 3+1 decomposition within the Hamiltonian framework. However, this approach presents a significant challenge when applied to Klein space due to its two equivalent temporal dimensions. The absence of a preferred time direction means that selecting either temporal coordinate would explicitly break the symmetry between them, leading to an ambiguous quantization procedure. To resolve this issue, we introduce a novel quantization scheme that preserves the inherent symmetry of Klein space while maintaining the essential features of canonical quantization.  Instead of using one of two temporal coordinates, we adopt the temporal "radial coordinate" $q$—the invariant length in the time plane-to foliate the space.  In this framework, we employ the coordinates specified in \eqref{eq:polar} to perform the canonical quantization.
The induced metric of $q = $ constant hypersurface $\Sigma_q$ is then 
\begin{equation}
    \td s^2_{\Sigma_q} = - q^2 \td \psi^2 + \td \vec x^2, \quad \vec x = (x^2, x^3).
\end{equation}
Therefore, the determinant of the induced metric is $h = \text{det} [h_{ab}] = - q^2$. 

In this section, we investigate the canonical quantization for real scalar fields, with the Lagrangian expressed in terms of the coordinates in \eqref{eq:polar} as
\begin{equation}
    \begin{split}
        \mathcal{L} & = - \frac12 (\p_{\mu} \phi \p^{\mu} \phi + m^2 \phi^2) - V (\phi) \\
        & = \frac12 \left[ (\p_q \phi)^2 + \frac{1}{q^2} (\p_{\psi} \phi)^2 - (\vec \p \phi)^2 - m^2 \phi^2 \right] - V (\phi),
    \end{split}
\end{equation}
where $V (\phi)$ denotes the interacting term  relying only on the field $\phi$. The conjugate momentum is then
\begin{equation}
    \pi_q = \sqrt{h} \frac{\p \mathcal{L}}{\p (\p_q \phi)} = i q \p_q \phi.
\end{equation}
%\hzz{sout:(We focus only on the weak coupling case, where $V(\phi)$ is very small.) (I think our discussion in this paper does not rely on the weak coupling because we did not perform any perturbative calculation.) }
Therefore, the canonical quantization yields
\begin{equation} \label{eq:cancommu}
    \begin{split}
        [\phi (q, \psi_2, \vec x_2), q \p_q \phi (q, \psi_1, \vec x_1)] & = \delta (\psi_{12}) \delta^2 (\vec x_{12}), \\
        [\phi (q, \psi_1, \vec x_1), \phi (q, \psi_2, \vec x_2)] & = 0,
    \end{split}
\end{equation}
where
\begin{equation}
    \delta (\psi_{12}) = \frac{1}{2 \pi} \sum_{n \in \mathbb Z} e^{i n \psi_{12}}, \qquad \int^{2 \pi}_0 \td \psi \delta(\psi - \psi') f (\psi) = f (\psi').
\end{equation}
%\cb{(After introducing the mode expansions, the canonical quantization relation \eqref{eq:cancommu} can be rewritten in terms of annihilation or creation operators, which proportional to the Klein-Gordon inner product through the field $\phi$ and the basis $B$)?}\hzz{(The relevant discussions are near Eq.(3.26). I think here is just a definition of Klein-Gordon product in hypersurfaces $\Sigma_q$.)}\cb{(Please correct the statement.)}
%\mxc{maybe we can use this statement: 

Before moving on, it is also useful to introduce the definition of Klein-Gordon (KG) inner product between two fields $\phi^1$ and $\phi^2$ %(or maybe place it at where the Klein inner product first appear, such as \eqref{eq:ePW0}, or \eqref{eq:ap=anJN'}?)}
\begin{equation} \label{eq:Kleininner}
    \< \phi^1, \phi^2 \> \equiv i \int_{\Sigma_q} \td \psi \td^2 x (\phi^2 \pi^1_q - \phi^1 \pi_q^2) = \int_{\Sigma_q} \td \psi \td^2 x \phi^1 \overset{\longleftrightarrow}{q \p_q} \phi^2.
    %= \int_{\Sigma_q} q \td \psi \td^2 x (\phi_2 \p_q \phi_1 - \phi_1 \p_q \phi_2).
\end{equation}
Here we need only partial derivative $\p_q$ rather than covariant derivative,  as we are considering scalar fields.

\subsection{Free real scalar field}

\subsubsection{Mode expansion}

Let us first consider the free real scalar ($V (\phi) = 0$), with the equation of motion (E.o.M)
\begin{equation} \label{eq:EOMq}
    (-\p^2 + m^2) \phi = \left[ \frac{1}{q} \p_q (q \p_q) + \frac{1}{q^2} \p^2_{\psi} + (m^2 - \vec \p^2) \right] \phi (q, \psi, \vec x) = 0\,.
\end{equation}
%Unfortunately, the dependence of $a_{\theta,\vec{p}}^{(J)} \,, a_{\theta,\vec{p}}^{(J) *}$ results in a situation in which the two equations of canonical quantization \eqref{eq:cancommu} cannot be satisfied simultaneously. Consequently, it can be predicted that some additional modes, independent with plane waves, will emerge at the quantum level, while these extra modes do not manifest at the classical level. Fortunately, the extra mode can be easily introduced by using the coordinates in \eqref{eq:polar}. In this sense, the E.o.M can be rewritten as
The solution can be generally expressed as
\begin{equation}
    \phi (q, \psi, \vec x) = \sum_{n \in \mathbb Z} e^{- i n \psi} \int \td^2 \vec p e^{i \vec p \cdot \vec x} f_n (\omega_{\vec p} q) a_{n, \vec p},
\end{equation}
where the function $f_n (q)$ obeys the equation
\begin{equation} \label{eq:Besseleq}
    u^2 f''_n (u) + u f'_n (u) + (u^2 - n^2) f_n (u) = 0, \qquad u=\omega_{\vec p} q.
\end{equation}
This is the Bessel equation, which has two linearly independent solutions
\begin{equation}
    \begin{split}
        J_n (x) & = \sum_{k = 0}^{\infty} \frac{(-)^k (x/2)^{n+2k}}{k! \Gamma (n+k+1)} = (-)^n J_{-n} (x), \\
        N_n (x) & = \lim_{\nu \rightarrow n} \frac{J_{\nu} (x) \cos \pi \nu - J_{- \nu} (x)}{\sin \pi \nu},
    \end{split}
\end{equation}
where $J_n, N_n$ are referred to as $n$-th order Bessel function and Neumann function (also called Bessel function of the second kind), and they form the Bessel basis as the solutions of the Klein Gordon equation when multiplied with $e^{-i n\psi+i \vec{p}\cdot\vec{x}}$. However, based on the following asymptotic behavior of $N_n, J_n$
\begin{equation} \label{eq:asympJNH}
    \begin{split}
        & x \rightarrow 0: \ 
        \begin{cases}
            & J_0 (x) \rightarrow 1, \quad J_{n \neq 0} \rightarrow \frac{1}{n!} \left( \frac x2 \right)^n, \\
            & N_0 (x) \sim \frac{2}{\pi} \ln x, \quad N_{n \neq 0} (x) \rightarrow - \frac{(n-1)!}{\pi} \left( \frac2x \right)^n
        \end{cases} \\
        & x \rightarrow \infty: \ 
        \begin{cases}
            & J_n (x) \rightarrow \sqrt{\frac{2}{\pi x}} \cos \left( x - \frac n2 \pi - \frac{\pi}{4} \right) \\
            & N_n (x) \rightarrow \sqrt{\frac{2}{\pi x}} \sin \left( x - \frac n2 \pi - \frac{\pi}{4} \right)
        \end{cases},
    \end{split}
\end{equation}
the Neumann function $N_n (\omega_{\vec p} q)$ is divergent when $q \rightarrow 0$. Therefore, the $N_n$ modes are forbidden in the classical solution, as the field $\phi (x)$ must remain finite for any value of the coordinates at the classical level. As a result, the classical solution is given by
\begin{equation} \label{eq:PWmodeJ}
    \phi^{\text{cl}} (q,\psi,\vec{x}) = \sqrt{\frac{\pi}{2}} \sum_{n \in \mathbb Z} \int \frac{\td^2 \vec p}{(2 \pi)^3} a^{(J)}_{n, \vec p} J_n ( \omega_{\vec p} q ) e^{i (\vec p \cdot \vec x - n \psi)},
\end{equation}
where the superscript ``cl" is the abbreviation of ``classical", and $a^{(J)}$ is independent of the coordinates $(q, \psi, \vec x)$. The reality condition $\phi^* = \phi$ is then equivalent to
\begin{equation} \label{eq:realcl}
    a_{n, \vec p}^{(J) *} = (-)^n a_{-n, -\vec p}^{(J)}\,,
\end{equation}
because of $J_n = (-)^n J_{-n}$. The classical solution for general case $\mathbb R^{n, m} (n, m \geq 2)$ will be discussed in \cite{Chen:2025eeh}.

The mode expansion (\ref{eq:PWmodeJ}) indicates that there are no two linearly independent modes for the classical solution in Klein space. This becomes much clearer when compared to the Minkowski case. To facilitate this comparison, we re-express the classical solution \eqref{eq:PWmodeJ} in terms of plane-wave modes using Cartesian coordinates as
\begin{align}
    \phi^{\text{cl}} (q,\psi,\vec{x})& = \frac{1}{2} \int \frac{\td^4 p}{(2 \pi)^3} \delta(p^2+m^2)  a_p^{(J)} e^{i p x}\\
    & = \frac{1}{4} \int \frac{\td \theta \td^2 \vec{p}}{(2 \pi)^3}  a_{\theta,\vec{p}}^{(J)} e^{i(-\omega_{\vec{p}} q \cos(\theta-\psi)+\vec{p} \cdot\vec{x})}  \notag\,,
\end{align}
where $\theta$ is the polar angle of the momentum component $p^0, p^1$, and we used the integral representation of Bessel function
\begin{equation} \label{eq:eth=J}
    J_n(\omega_{\vec{p}} q) = \frac{i^n}{2\pi} \int_{-\pi}^{\pi} \td \theta e^{- i n\theta}e^{-i \omega_{\vec{p}}q \cos \theta}\, ,
\end{equation}
and the relation between the coefficients
\begin{equation} \label{eq:athpaJ}
    a_{\theta, \vec p}^{(J)} = a_p^{(J)} \Big|_{p^2 + m^2 = 0} = \sqrt{\frac{2}{\pi}} \sum_{n \in \mathbb Z} i^n a_{n,\vec{p}}^{(J)} e^{i n\theta}\,,
     %a_{n,\vec{p}}^{(J)} =i^n\int \td\theta e^{-i n\theta}a_{\theta,\vec{p}}\, .
\end{equation}
In comparison, the mode expansions via plane waves in Minkowski
\begin{equation}
    \phi^{\text{cl}}_{\text{Mink}}(t,\vec x)=\int\frac{\td^3 \vec p}{(2\pi)^3 2\omega_{\vec p}}\left[a_{\vec p}e^{-i \omega_{\vec p} t+i\vec p\cdot\vec x}+a^*_{-\vec{p}}e^{i \omega_{\vec p}t+i \vec p\cdot\vec x} \right]\,, 
\end{equation}
do have two linearly independent modes $e^{-i \omega_{\vec p} t+i\vec p\cdot\vec x}$ and $e^{i \omega_{\vec p}t+i \vec p\cdot\vec x}$.
%Therefore, the coefficients of the Bessel modes $a^{(J)*}, a^{(J)}$ correspond to the plane wave modes $a^\dagger, a$ in the Minkowski case, where $a^\dagger$ and $a$ are independent coefficients. However, in the Klein space, 
%The reality condition \eqref{eq:realcl} then leads to mutual dependence between $a^{(J)*}$ and $a^{(J)}$ %\cb{mutual dependence between operators $a^{(J)*}$ and $a^{(J)}$ for plane waves in Klein space (We did't define the vacuum and couldn't classify which are creation operators)}
%\begin{equation} \label{eq:realclPW}
    %a_{\theta, \vec p}^{(J)} = a_{\theta + \pi, - \vec p}^{(J) *}\,.
%\end{equation}
%the KG product between the plane wave bases $e^{\pm i p \cdot x}$ vanishes, since they are on-shell. %Due to the dependence of the plane wave modes, the classical free fields $\phi$ are not independent with their conjugation $\pi_q = i q \p_q \phi$. As a result, it is not possible to satisfy the two requirements of \eqref{eq:cancommu} simultaneously when dealing with the canonical quantization. 
%Consequently, there are no two independent plane wave modes for the classical solution in Klein space, which exhibits completely different properties compared to the case in Minkowski spacetime.
In Minkowski spacetime, the creation and annihilation operators can create plane wave modes at the past infinity and the future infinity, respectively. Thus, there must be two independent modes associated with two different conformal boundaries. But in Klein space, whose Penrose diagram is topologically equivalent to gluing together the two conformal boundaries of the Penrose diagram of Minkowski spacetime, there is only one conformal boundary. The gluing of the boundaries transforms the worldlines of massive plane waves to closed curves (see the black dashed line in Fig. \ref{fig:KleinPen})
%\footnote{\cb{\sout{Actually, for the worldline of massless plane wave (the yellow wavy line in Fig. \ref{fig:KleinPen}), the effective length of its two end points vanishes, since the end points are on the null infinity. Therefore, the worldline of the massless plane wave can also be regarded as a closed curve.}(The statement here is hand-waving).}}
, which means that the outgoing and incoming processes of plane waves on the single conformal boundary are no longer two independent processes. Instead, they are identical to each other.
%Therefore, coefficients $a_{\theta + \pi, - \vec p}^{(J)}$ and $a_{\theta, \vec p}^{(J)}$, corresponding to the outgoing and incoming plane waves, are related due to the symmetry and interchangeability of the outgoing and incoming processes on the single conformal boundary.}

%As shown above, there do not exist two independent modes for the classical solution in Klein space. %as in the Minkowski spacetime.
However, only one linearly independent mode causes some severe problems. One is that the classical free fields $\phi$ are not independent of their conjugation $\pi_q = i q \p_q \phi$. This can be shown by using the KG inner product.
%On one hand, this can be observed from the perspective of the KG inner product.
Note that the dual of the Klein-Gordon current $J^{12} := \phi^1 \overleftrightarrow{\p_\mu} \phi^2 \td x^\mu$ constructed by the on-shell fields $\phi^1, \phi^2$ is closed
\begin{align}
    \td*J^{12} = - \p^{\mu} \left( \phi^1 \overleftrightarrow{\p_\mu} \phi^2 \right) \td^4 x = m^2 \phi^1 \phi^2 - m^2 \phi^1 \phi^2 = 0,
\end{align}
which has also been demonstrated in \cite{Melton:2024pre} for massless case. 
%Here we have proved the massive free case with the on-shell condition $\p^{\mu} \p_{\mu} \phi^i = m^2 \phi^i (i = 1, 2)$. 
This results in the vanishing of the KG inner product between two fields
\begin{align} \label{eq:int*J=0}
    \< \phi^1, \phi^2 \> \propto \int_{\Sigma_{q_0}}*J^{12} = \int_{M_4} \td*J^{12}=0.
\end{align}
Here, the hypersurfaces $\Sigma_{q_0}$ are always the boundary $\partial M_4$ of a four-dimensional region $M_4\equiv\{(q,\psi,\vec{x})| 0 \leq q \leq q_0\}$. In particular, provided that the field $\phi$ is on-shell, the conjugate momentum $\pi_q = i q \p_q \phi$ also satisfies the Klein-Gordon equation, indicating the vanishing of the inner product 
\begin{equation}
    \<\phi^{\text{cl}}, \pi_q\> = 0,
\end{equation}
The vanishing of the inner product indicates the linear dependence of $\phi$ and $\pi_q$. As a result, it is not possible to satisfy the two relations in \eqref{eq:cancommu} simultaneously when dealing with the canonical quantization.

The other more technical problem is that when we consider amplitudes in flat space, we need to use the KG inner product
\begin{equation} \label{eq:ePW0}
    \<e^{\pm i p x}, \phi^{\text{cl}} (x) \> = 0\,,
\end{equation}
to define an asymptotic scattering state, see (\ref{eq:AsymScattStat}) and (\ref{eq:aInScattering}) later. But here it just vanishes due to the same reason as in the above discussions. 

%Furthermore, compared with the case in Minkowski space, the KG inner product of the field and the plane wave in the Klein space can no longer determine the coefficients of the plane wave modes, since the plane wave itself is on-shell
%\cb{(As I mentioned before, there is short of logic here: why we stick to plane wave, which only appears in the mode expansion in the Cartesian coordinates? If the plane wave is important, we should elaborate.)}
%\cb{The superscript "PW" denotes the plane wave mode expansion of the field, in contrast to the Bessel function expansion  that we will introduce subsequently, while the meaning of superscript ``$(J)$" will be clarified later. Actually, unlike the theories in Minkowski space, the operators $a_{\theta,\vec{p}}^{(J)} \,, a_{\theta,\vec{p}}^{(J) *}$ in Klein space are no longer independent of each other.} The following lies some reasons:\cb{(I feel the following points are short of logic. If we use $q$ as the time coordinate, why we discuss the plane-wave expansion in (3.7)? It is natural to study (3.12) directly, right?)} 

To overcome these problems, it is essential to introduce additional modes besides the ones relating to Bessel function. These additional modes could emerge at the quantum level, while they are forbidden at the classical level. {The Neumann modes $N_n$, previously discarded by the classical regular condition at $q = 0$, are appropriate candidates. They can be retrieved by removing the original point in the Klein space to obtain a new manifold $\mathbb{K}^{2,2}-\{0\}$, where the quantum fields are defined. Then, the Klein-Gordon inner product between on-shell fields will no longer vanish, because the $\Sigma_{q_0}$ in \eqref{eq:int*J=0} is not the only boundary of $M_4 - \{0\} \equiv \{(q,\psi,\vec{x})| 0 < q \leq q_0\}$. %which introduces an extra boundary term $\int_{\Sigma_{q_0 = 0}} *J^{12}$ after implementing the Stokes formula. 
Consequently, $\phi$ and $\pi_q$ are independent after introducing $N_n$. Furthermore, the classical requirement for the field $\phi$ to be regular at $q = 0$ is modified in the quantum level:  we only require its regularity when acting on the vacuum state $\ket{0}$
\begin{equation} \label{eq:regquant}
    \phi (q, \psi, \vec x) \ket{0} \Big|_{q = 0} = \text{finite}.
\end{equation}
%This is analogous to the definition of asymptotic states in CFT \cite{DiFrancesco:1997nk}. 
A closely related issue also arises in the discussion of radial quantization in CFT \cite{DiFrancesco:1997nk}, where operators are usually expanded in Laurent modes, which includes many divergent modes at the origin. Instead of dropping these divergent modes, we also retain them and require their action on the vacuum state to be regular by requiring the divergent modes annihilate the vacuum. This gives the definition of the vacuum state and the well-defined primary states in CFT.

As a result, the mode expansion of free scalar field should be reformulated as
\begin{equation} \label{eq:JNmode}
    \phi (q, \psi, \vec x)  = \sqrt{\frac{\pi}{2}} \sum_{n \in \mathbb Z} \int \frac{\td^2 \vec p}{(2 \pi)^3} \left[ a^{(J)}_{n, \vec p} J_n ( \omega_{\vec p} q ) + a^{(N)}_{n, \vec p} N_n ( \omega_{\vec p} q ) \right] e^{i (\vec p \cdot \vec x - n \psi)} \,.
\end{equation}
Here, $a^{(J)}$ and $a^{(N)}$ are linearly independent operators. %, and they are independent with the coordinates $(q, \psi, \vec x)$. 
The reality condition is then given by
\begin{equation} \label{eq:realaJN}
    a^{(J) *}_{n, \vec{p}} = (-)^n a^{(J)}_{- n,- \vec{p}}\, , \quad a^{(N) *}_{n, \vec{p}} = (-)^n a^{(N)}_{-n, - \vec{p}}\,,
\end{equation}
because of $N_n = (-)^n N_{-n}, \ J_n = (-)^n J_{-n}$. Similar to the discussions in CFT, the regular condition \eqref{eq:regquant} is explicitly given by annihilating the divergent modes when acting on the vacuum state
\begin{equation} \label{eq:regular}
    a^{(N)}_{n, \vec p} \ket{0} = 0 \quad (\forall n, \vec p),
\end{equation}
as the $q = 0$ divergences come from $N_n$ modes. Since the state $|0\>$ is annihilated by the Neumann function coefficient, we will call it the Neumann vacuum, in contrast with the Hankel vacuum $\<\infty|$ which will be defined in (\ref{eq:InftyStateAnnihilation}) later. Now the Klein-Gordon product between the plane-wave and the field is not vanishing
\begin{equation} \label{eq:ap=anJN'}
    \begin{split}
        \< e^{- i p \cdot x}, \phi (x) \> & = \sqrt{\frac{\pi}{2}} \sum_{n \in \mathbb Z} i^n e^{i n \theta} a^{(N)}_{n, \vec p} J_n (\omega_{\vec p} q) \overset{\longleftrightarrow}{q\p_q} N_n (\omega_{\vec p} q) %\\
        %& \equiv \sqrt{\frac{2}{\pi}} \sum_{n \in \mathbb Z} i^n e^{i n \theta} a^{(N)}_{n, \vec p},
    \end{split}
\end{equation}
The above equation is independent of $q$, due to the following identity\footnote{\label{fn:JqpqN}This formula can be verified in two steps. First, it can be proved directly at the asymptotic region $q \rightarrow 0, \infty$ by using the asymptotic behavior \eqref{eq:asympJNH}. Second, the derivative of $J_n (\omega_{\vec p} q) \overset{\longleftrightarrow}{q \p_q} N_n (\omega_{\vec p} q)$ by $q$ vanishes because $J_n (\omega_{\vec p} q), N_n (\omega_{\vec p} q)$ satisfies the Bessel equation \eqref{eq:Besseleq}. Therefore this relation holds for any $q$.}
\begin{equation} \label{eq:asymJN}
    J_n (\omega_{\vec p} q) \overset{\longleftrightarrow}{q \p_q} N_n (\omega_{\vec p} q) =\frac{2}{\pi} \qquad (\forall q \geq 0).
\end{equation}
%which can be verified using \eqref{eq:asympJNH} and the Bessel equation. 
Then, one obtains 
\begin{equation} \label{eq:aN=KG0}
    a^{(N)}_{\theta, \vec p} \equiv \sqrt{\frac{2}{\pi}} \sum_{n \in \mathbb Z} i^n e^{i n \theta} a^{(N)}_{n, \vec p} = \< e^{- i p \cdot x}, \phi (x) \>.
\end{equation}
%Similar to the interpretation in Minkowski space, the operator $a^{(N)}_{\theta, \vec p}$ here can be interpreted as the operators generating the plane waves at the boundaries $q \rightarrow 0, \infty$. Similarly, the operators $a^{(J)}_{n,\vec{p}}, a^{(N)}_{n,\vec{p}}$ can be derived through Klein-Gordon inner product \eqref{eq:Kleininner} between the field and the Bessel basis. %For comparison, we will also express $a^{(N)}$ in the following. 
For convenience, one can perform Fourier transformation to Eq.(\ref{eq:aN=KG0}) and obtains
\begin{equation}\label{eq:aN}
    \begin{split}
        a_{n, \vec p}^{(N)} & = \sqrt{\frac{\pi}{2}}\left\< e^{-i (\vec p \cdot \vec x - n \psi)} J_n (\omega_{\vec p} q), \phi (q, \psi, \vec x) \right\> \\
        & = - \sqrt{\frac{\pi}{2}} \int \frac{\td^2 \vec x \td \psi}{(2 \pi)^3} e^{-i (\vec p \cdot \vec x - n \psi)} \phi (q, \psi, \vec x) \overset{\longleftrightarrow}{q \p_q} J_n (\omega_{\vec p} q).
    \end{split}
\end{equation}
Similarly, the $a_{n, \vec p}^{(J)}$ can also be derived in terms of the Klein-Gordon inner product as
\begin{equation}\label{eq:aJ}
    \begin{split}
        a_{n, \vec p}^{(J)} & = - \sqrt{\frac{\pi}{2}}\left\< e^{-i (\vec p \cdot \vec x - n \psi)} N_n (\omega_{\vec p} q), \phi (q, \psi, \vec x) \right\> \\
        & = \sqrt{\frac{\pi}{2}} \int \frac{\td^2 \vec x \td \psi}{(2 \pi)^3} e^{-i (\vec p \cdot \vec x - n \psi)} \phi (q, \psi, \vec x) \overset{\longleftrightarrow}{q \p_q} N_n (\omega_{\vec p} q).
    \end{split}
\end{equation}
%\begin{equation} \label{eq:aJaN}
    %\begin{split}
        %\left\<e^{-i (\vec p \cdot \vec x - n \psi)} N_n (\omega_{\vec p} q),  \phi (q, \psi, \vec x) \right\> & 
        %\int \td^2 \vec x \td \psi e^{i (\vec p \cdot \vec x - n \psi)} \phi (q, \psi, \vec x) \overset{\longleftrightarrow}{q \p_q} N_n (\omega_{\vec p} q) &
        %= a_{n, \vec p}^{(J)} \omega_{\vec p} \sqrt{\frac{\pi}{2}} J_n (\omega_{\vec p} q) \overset{\longleftrightarrow}{q \p_q} N_n (\omega_{\vec p} q), \\
        %\left\< e^{-i (\vec p \cdot \vec x - n \psi)} J_n (\omega_{\vec p} q), \phi (q, \psi, \vec x) \right\> & %=\int \td^2 \vec x \td \psi e^{i (\vec p \cdot \vec x - n \psi)} \phi (q, \psi, \vec x) \overset{\longleftrightarrow}{q \p_q} J_n (\omega_{\vec p} q) &
        %= a_{n, \vec p}^{(N)} \omega_{\vec p} \sqrt{\frac{\pi}{2}} N_n (\omega_{\vec p} q) \overset{\longleftrightarrow}{q \p_q} J_n (\omega_{\vec p} q),
    %\end{split}
%\end{equation}
%\begin{equation}\label{eq:aJaN=phipi}
    %\begin{split}
        %a_{n, \vec p}^{(J)} \sim %\xlongequal{q \rightarrow 0,\infty} 
        %& \sqrt{\frac{\pi}{2}} \int \frac{\td^2 \vec x \td \psi}{(2 \pi)^3} e^{i (\vec p \cdot \vec x - n \psi)} \phi (q, \psi, \vec x) \overset{\longleftrightarrow}{q \p_q} N_n (\omega_{\vec p} q), \\
        %a_{n, \vec p}^{(N)} %\xlongequal{q \rightarrow 0,\infty} 
        %\sim & - \sqrt{\frac{\pi}{2}} \int \frac{\td^2 \vec x \td \psi}{(2 \pi)^3} e^{i (\vec p \cdot \vec x - n \psi)} \phi (q, \psi, \vec x) \overset{\longleftrightarrow}{q \p_q} J_n (\omega_{\vec p} q)\,,
    %\end{split}
%\end{equation}
It is important to reiterate that the Eqs \eqref{eq:aJ} and \eqref{eq:aN} hold for any $q$, because their left-hand sides are all independent of $q$. The canonical quantization \eqref{eq:cancommu} is thus equivalent to
\begin{equation}\label{eq:aJaNCom}
    \left[ a^{(J)}_{n, \vec p_1}, a^{(N)}_{m, \vec p_2} \right] = (-)^n (2 \pi)^3 \delta_{m+n, 0} \delta^2 (\vec p_1 + \vec p_2) %\frac{\pi}{2} \omega_{\vec p} J_n (\omega_{\vec p} q) \overset{\longleftrightarrow}{q \p_q} N_n (\omega_{\vec p} q),
\end{equation}
or equivalently,
\begin{equation}
    \left[ a^{(J)}_{n, \vec p_1}, a^{(N) *}_{m, \vec p_2} \right] = (2 \pi)^3 \delta_{m, n} \delta^2 (\vec p_{12}).
\end{equation}

\subsubsection{Hamiltonian and evolution operator} \label{sec:H0U0}

The Hamiltonian of the free real scalar can be derived from the Legendre transformation of the Lagrangian
\begin{equation}
    \begin{split}
        H_q^{(0)} & = \int q \td \psi \td^2 \vec x \left( \frac{1}{\sqrt{h}} \pi_q \p_q \phi - \mathcal{L} \right) \\
        & = \frac 12 \int \td \psi \td^2 \vec x \left[ - \frac{1}{q} \pi_q^2 - \frac{1}{q} (\p_\psi \phi)^2 + (\vec \p \phi)^2 + m^2 \phi^2 \right].
        %& = \frac q2 \int \td \psi \td^2 \vec x \left[ (\p_q \phi)^2 - \frac{1}{q^2} (\p_\psi \phi)^2 + (\vec \p \phi)^2 + m^2 \phi^2 \right]
        %& = \frac q2 \int \td \psi \td^2 \vec x \left[ (\p_q \phi)^2 + \phi \left( m^2 - \vec \p^2 + \frac{1}{q^2} \p_\psi^2 \right) \phi \right] \\
        %& = \frac 12 \int \td \psi \td^2 \vec x \left[ - \frac{1}{q} (\pi_q)^2 + i \phi \p_q \pi_q \right],
    \end{split}
\end{equation}
%where we used the integral by parts and the E.o.M. 
The Hamiltonian is then $q$-dependent. Furthermore, the Hamiltonians at different values of $q$ do not commute with each other, while Hamiltonians at the same $q$ commute
\begin{equation}
    \left[ H^{(0)}_{q_1}, H^{(0)}_{q_2} \right] \neq 0, \quad \left[ H^{(0)}_q, H^{(0)}_q \right] = 0.
\end{equation}
The commutators between $H^{(0)}_q$ and the field $\phi$ or the conjugation momentum $\pi_q$ at the same $q$ are derived from the canonical commutator \eqref{eq:cancommu} as
\begin{equation}
    \begin{split}
        \left[ H^{(0)}_q, \phi (q, \psi, \vec x) \right] & = - \p_q \phi (q, \psi, \vec x), \\
        \left[ H^{(0)}_q, \pi_q (q, \psi, \vec x) \right] & = - \p_q \pi_q (q, \psi, \vec x),
    \end{split}
\end{equation}
which are the Heisenberg equations. The solution of the Heisenberg equations are
\begin{equation}
    \begin{split}
        \phi (q,\psi, \vec x) & = U^{-1}_{(0)} (q, q_0) \phi (q_0, \psi, \vec x) U_{(0)} (q, q_0)\,, \\
        \pi_q (q,\psi, \vec x) & = U^{-1}_{(0)} (q, q_0) \pi_{q_0} (q_0, \psi, \vec x) U_{(0)} (q, q_0)\,.
    \end{split}
\end{equation}
Here, $q_0$ is the reference time (or initial time), and the evolution operator $U_{(0)} (q, q_0)$ with the initial condition $U_{(0)} (q_0, q_0) = 1$ satisfies the schr$\Ddot{\text o}$dinger equation
\begin{equation} \label{eq:Heiseq}
    \p_q U_{(0)} (q, q_0) = U_{(0)} (q, q_0) H_q^{(0)}, \quad \p_q U^{-1}_{(0)} (q, q_0) = - H_q^{(0)} U^{-1}_{(0)} (q, q_0).
\end{equation}
These two equations are actually equivalent, and they have the solution 
\begin{equation}
    U_{(0)} (q, q_0) = \mathcal{Q}^{-1} \text{exp} \left[\int^{q}_{q_0} H^{(0)}_{q'} \td q' \right] %=\mathcal{Q} \text{exp} \left[\int^{q}_{q'} H^{(0)(S)}_q \td q \right]
    \quad\text{when}\ q>q_0\,,
\end{equation}
and
\begin{equation}
    U_{(0)} (q,q') = U^{-1}_{(0)}(q',q)\quad\text{when}\ q<q'\,,
\end{equation}
where $\mathcal{Q}^{-1}$ is the anti-$q$ order operator placing the operators with larger $q$ on the right. Note that $H_q^{(0)}$ is actually the Hamiltonian in the Heisenberg picture. The evolution operator $U_{(0)} (q, q_0)$ takes a complex form because the fact that  $H_q^{(0)}$'s at different $q$ do not commute with each other. %the non-commutative property of . %However, this is not similar to the free scalar in the Minkowski case, where the evolution operator $U_{Min} (q, q_0) = e^{i H^{Min} (t - t_0)}$ typically does not introduce time-ordered products. %because $H^{Min}$ for free theory is time-independent. 
%Fortunately, $U(q, q_0)$ can be alternatively expressed in the similar way as that in the Minkowski case with the help of schr$\Ddot{\text o}$dinger picture.

The Hamiltonian in the Schr$\Ddot{\text o}$dinger picture is
\begin{equation}
    \begin{split}
        H_q^{(0) (S)} & = U_{(0)} (q, q_0) H_q^{(0)} U^{-1}_{(0)} (q, q_0) \\
        & = \frac 12 \int \td \psi \td^2 \vec x \left[ - \frac{1}{q} \pi_{q_0}^2 - \frac{1}{q} (\p_\psi \phi (q_0))^2 + (\vec \p \phi (q_0))^2 + m^2 (\phi (q_0))^2 \right],
    \end{split}
\end{equation}
where $\phi (q_0)$ is a shorthand version of $\phi (q_0, \psi, \vec x)$. The Hamitonian in these two pictures are linked at the initial time $q_0$
\begin{equation}
    H^{(0) (S)}_{q_0} = H^{(0)}_{q_0}.
\end{equation}
Note that even in the Schr$\Ddot{\text o}$dinger picture, the Hamiltonian for free scalar still depends on $q$, and 
\begin{equation}
    \left[ H^{(0) (S)}_{q_1}, H^{(0) (S)}_{q_2} \right] \neq 0 \quad (q_1\neq q_2),\quad\quad  \left[ H^{(0) (S)}_q, H^{(0) (S)}_q \right] = 0.
\end{equation}
The Heisenberg equation \eqref{eq:Heiseq} can be rewritten in term of $H^{(0) (S)}_q$ as
\begin{equation}
    \p_q U_{(0)} (q, q_0) = H^{(0) (S)}_q U_{(0)} (q, q_0).
\end{equation}
%where $\mathcal{Q}$ represents the $q$-order operator placing the fields with larger $q$ to the left of those with smaller $q$ and $\mathcal{Q}^{-1}$ is the anti-$q$ order operator placing the operators with larger $q$ on the right. 
%Here $H_q^{(0)(S)}$ is the Hamiltonian in the Schrödinger picture related to the Hamiltonian in the Heisenberg picture $H^{(0)}_q$ as
%\begin{equation}
    %H^{(0)}_q =U^{-1}(q,0)H_q^{(0)(S)}U(q,0)\,.
%\end{equation}
Then, the evolution operator can be expressed in terms of Hamiltonian in the Schr$\Ddot{\text o}$dinger picture as
\begin{equation}
    U_{(0)} (q, q_0) = \mathcal{Q} \text{exp} \left[\int^{q}_{q_0} H^{(0) (S)}_{q'} \td q' \right], \quad \text{when} \ q> q_0
\end{equation}
where $\mathcal{Q}$ represents the $q$-ordering operator placing the fields with larger $q$ to the left of those with smaller $q$.

\subsubsection{Correlation functions}

On the other hand, the quantum field theory in Klein space can also be regarded as the analytical continuation from Euclidean space by implementing Wick rotation $q_E \rightarrow i q (1- i \epsilon)$. Here, $q_E$ denotes the radial direction on 2D plane $(x^0_E, x^1_E)$ in Euclidean space parametrized by the coordinates $(x^0_E, x^1_E, x^2, x^3)$. This is similar to the analytical continuation from Euclidean space to Minkowski space $x^0_E \rightarrow i x^0 (1 - i \epsilon)$, which leads to the $x^0$-ordered (time-ordered) correlation function. For more information about Wick rotation, see Appendix \ref{ap:analyCon}. Therefore, the $n$-point correlation function for QFT in Klein space is similarly defined by the $q$-ordered expectation of $n$ operators, expressed as
\begin{equation}
    \< \phi_1 (x_1) \cdots \phi_n (x_n) \> = \bra{\infty} \mathcal{Q} \phi_1 (x_1) \cdots \phi_n (x_n) \ket{0}\,.
\end{equation}
Instead of using $\bra{0}$, %which is the Hermitian conjugation of $|0\>$, 
we adopt the state $\bra{\infty}$ as the bra vacuum state in the correlation function, because here we need to work in an ``in-out" formalism rather than an ``in-in" formalism. The ``in-out" formalism can be translated to the usual path integral and related to the results from analytical continuation, while the ``in-in" formalism can be translated to a closed-time-path functional integral \cite{Schwinger:1960qe}%(Schwinger 1961). 
. The field $\phi$ at the asymptotic infinity behaves as
\begin{equation}
    \begin{aligned}
        \phi (q, \psi, \vec x) % &= \sqrt{\frac{\pi}{2}} \sum_{n \in \mathbb Z} \int \frac{\td^2 \vec p}{(2 \pi)^3} \left[ a^{(J)}_{n, \vec p} J_n ( \omega_{\vec p} q ) + a^{(N)}_{n, \vec p} N_n ( \omega_{\vec p} q ) \right] e^{i (\vec p \cdot \vec x - n \psi)} \\
        &=\sqrt{\frac{\pi}{2}} \sum_{n \in \mathbb Z} \int \frac{\td^2 \vec p}{(2 \pi)^3} \left[ a^{(H^{(1)})}_{n, \vec p} H_n^{(1)} ( \omega_{\vec p} q ) + a^{(H^{(2)})}_{n, \vec p} H_n^{(2)} ( \omega_{\vec p} q ) \right] e^{i (\vec p \cdot \vec x - n \psi)}\\
        %&\rightarrow \sqrt{\frac{\pi}{2}} \sum_{n \in \mathbb Z} \int \frac{\td^2 \vec p}{(2 \pi)^3} \sqrt{\frac{2}{\pi \omega_{\vec p}q}}\left[ a^{(J)}_{n, \vec p} \cos(\omega_{\vec p}q-\frac{n\pi}{2}-\frac{\pi}{4}) + a^{(N)}_{n, \vec p} \sin(\omega_{\vec p}q-\frac{n\pi}{2}-\frac{\pi}{4}) \right] e^{i (\vec p \cdot \vec x - n \psi)}\\
        & \underset{q \to \infty}{\longrightarrow} \sqrt{\frac{\pi}{2}} \sum_{n \in \mathbb Z} \int \frac{\td^2 \vec p}{(2 \pi)^3} \sqrt{\frac{2}{\pi \omega_{\vec p}q}}\left[ a^{(H^{(1)})}_{n, \vec p} e^{i\left(\omega_{\vec p}q-\frac{n\pi}{2}-\frac{\pi}{4}\right)} + a^{(H^{(2)})}_{n, \vec p} e^{-i\left(\omega_{\vec p}q-\frac{n\pi}{2}-\frac{\pi}{4}\right)} \right] e^{i (\vec p \cdot \vec x - n \psi)}
    \end{aligned}
\end{equation}
with
\begin{equation}
    a^{(H^{(1)})}_{n, \vec p}=\frac{a^{(J)}_{n, \vec p}-i a^{(N)}_{n, \vec p}}{2}\,,\quad a^{(H^{(2)})}_{n, \vec p}=\frac{a^{(J)}_{n, \vec p}+i a^{(N)}_{n, \vec p}}{2}\,,
\end{equation}
where we reformulated the mode expansion \eqref{eq:JNmode} by the first-kind and the second kind Hankel functions $H^{(1)}_n (x)$ and $ H^{(2)}_n (x)$ defined as
\begin{equation}
    H_n^{(1)}(x)=J_n(x)+i N_n(x)\,,\quad H_n^{(2)}(x)=J_n(x)-i N_n(x)\,.
\end{equation}
The modes $e^{i\left(\omega_{\vec p}q-\frac{n\pi}{2}-\frac{\pi}{4}\right)}$ and $e^{-i\left(\omega_{\vec p}q-\frac{n\pi}{2}-\frac{\pi}{4}\right)}$ are similar to $e^{i\omega t}$ and $e^{-i\omega t}$ in the Minkowski case. Thus, one of them should be interpreted as the annihilation operator and the other as the creation operator concerning the state $\<\infty|$. We will see that the right choice is
\begin{equation}\label{eq:InftyStateAnnihilation}
    \<\infty|a_{n,\vec p}^{(H^{(1)})}=0\,.
\end{equation}
Since the state $\<\infty|$ is annihilated by the Hankel function coefficient acting from the right side, we will call it the Hankel vacuum.
%Note that at the original point, concerning the state $|0\>$, the $a^{(N)}_n$ should be the annihilation operator due to the regularity condition (\ref{eq:regular}). Then the conditions (\ref{eq:regular}) and (\ref{eq:InftyStateAnnihilation}) would be key points in calculating correlation functions via Wick contraction.
Now we can calculate any correlation functions via the Wick contraction, based on the annihilation conditions of Neumann vacuum (\ref{eq:regular}) and the Hankel vacuum (\ref{eq:InftyStateAnnihilation}). Let us consider the two-point propagator $\Delta (x - x') =\langle \infty| \mathcal{Q} \phi(x)\phi(x')|0\rangle$ as an example. Set $q>q'$ without loss of generality, we have
\begin{equation} \label{eq:Wickpropagator}
    \begin{aligned}
        & \<\infty| \phi(x)\phi(x')|0\>/\<\infty|0\>\\
        =&-\frac{i\pi}{2}\sum_n \int \frac{\td^2 \vec p}{(2\pi)^3}e^{i\vec p\cdot (\vec x-\vec x') -i n (\psi-\psi')}H_n^{(2)}(\omega_{\vec p}q)J_{n}(\omega_{\vec p}q')\\
        =&-\frac{i\pi}{2}\int\frac{\td^2 \vec p}{(2\pi)^3} H_0^{(2)}\left(\omega_{\vec p}\Delta q\right)e^{i \vec p \cdot (\vec x-\vec x')}\,,
    \end{aligned}
\end{equation}
with
\begin{equation}
    \Delta q=\sqrt{q^2+q'^2-2qq'\cos(\psi-\psi')}\,,
\end{equation}
where we used $J_{-n}=(-)^n J_n$ and the addition theorem for Bessel functions.
Note that the inner product $\<\infty|0\>$ is non-vanishing\footnote{Since we define the state $|0\>$ and $\<\infty|$ by the annihilation conditions, there are ambiguities in normalizations, see Eq.(\ref{eq:DefVacu}) and Eq.(\ref{eq:DefInftyStat}). We can choose normalization constants properly so that we have the normalization $\<\infty|0\>=1$, see the discussions below Eq.(\ref{eq:VacuCorrFunc}).}. 

The two-point propagator (\ref{eq:Wickpropagator}) is manifestly invariant under spacetime transformations. As demonstrated in Appendix \ref{ap:propagator}, it can be expressed in a covariant form
\begin{equation} \label{Greenfucntion}
    \Delta(x-x')= - \int \frac{\td^4 p}{(2\pi)^4} \frac{e^{i p (x-x')}}{p^2+m^2-i\epsilon} \bra{\infty} 0 \> \,.
\end{equation}
%The expression (\ref{eq:Wickpropagator}) is manifestly invariant under spacetime translations. In general, the two-point propagator can be expressed in a covariant form. 
This is the Green's function of the Klein-Gordon equation. To see this, %This can be achieved checked by acting 
we can act Klein-Gordon operator $-\p^2 +m^2$ on the two-point function
\begin{equation}
     \begin{split}
         & \left(\frac{\p_q}{q}(q\p_q)+\frac{\p_\phi^2}{q^2}-\vec{\p}^2+m^2\right) \langle \infty| \mathcal{Q} \phi(x)\phi(x')|0\rangle \\
         & =\frac{\p_q}{q}\eta(q-q')  \bra{\infty} [q\p_q \phi(x),\phi(x')] \ket{0} \,,
     \end{split}
\end{equation}
where $\eta(x)$ denotes the step function. %, and we used the normalization $\bra{\infty} 0\> = 1$. 
This differential equation can be re-expressed by using \eqref{eq:cancommu}
\begin{equation}
    \begin{split}
        (-\p_x^2+m^2) \Delta (x - x') & = - \frac{1}{q}\delta(q-q')\delta(\psi-\psi')\delta^2(\vec{x}-\vec{x}') \bra{\infty} 0 \> \\
        & = - \delta^4(x-x') \bra{\infty} 0 \>\,.
    \end{split}
\end{equation}
%The solution of this equation is
%\begin{equation} \label{Greenfucntion}
    %\Delta(x-x')= - \int \frac{\td^4 p}{(2\pi)^4} \frac{e^{i p (x-x')}}{p^2+m^2-i\epsilon} \bra{\infty} 0 \> \, .
%\end{equation}
%The above expression matches the result \eqref{eq:Wickpropagator} derived from the Wick contraction, see details in the Appendix \ref{ap:propagator}. 
We can easily check that the propagator \eqref{Greenfucntion} is exactly the solution of the above differential equation. This also serves as the verification of the annihilation condition \eqref{eq:InftyStateAnnihilation} for the Hankel vacuum $\bra{\infty}$. 
%\cb{(The normalization condition is nontrivial, and we should clarify this point more clearly.)}\hzz{(see the footnote.)} \mxc{The $\<\infty|0\>$ is added back in eq (3.50)-(3.53).}

\subsection{The interacting real scalar field and LSZ reduction formula}

For the interacting real scalar, the potential $V (\phi)$ in the Lagrangian is non-vanishing, such that the E.o.M \eqref{eq:EOMq} should be modified to include potential terms. %\mxc{%Here we discuss the perturbative quantum field theory by taking the potential as a small perturbation.
In a QFT textbook, people usually use the interaction picture, which is convenient for perturbative calculations, see appendix \ref{ap:PertExpand}. However, in this subsection, we adopt the Heisenberg picture in which the mode expansion can be formally performed in a similar form to that of free field theory \eqref{eq:JNmode},
\begin{equation} \label{eq:JNmodeq}
    \phi (q, \psi, \vec x)  = \sqrt{\frac{\pi}{2}} \sum_{n \in \mathbb Z} \int \frac{\td^2 \vec p}{(2 \pi)^3} \left[ a^{(J)}_{n, \vec p} (q) J_n ( \omega_{\vec p} q ) + a^{(N)}_{n, \vec p} (q) N_n ( \omega_{\vec p} q ) \right] e^{i (\vec p \cdot \vec x - n \psi)} \,.
\end{equation}
But now the coefficients $a^{(J)}(q), a^{(N)}(q)$ are no longer independent of the coordinate $q$. As a result, they do not satisfy \eqref{eq:aJaNCom} except for $q\rightarrow\infty$ where the field tends to be free. As usual, the reality condition also yields
\begin{equation}
    a^{(J) *}_{n, \vec{p}} (q) = (-)^n a^{(J)}_{- n,- \vec{p}} (q) \, , \quad a^{(N) *}_{n, \vec{p}} (q) = (-)^n a^{(N)}_{-n, - \vec{p}} (q) \,.
\end{equation}
%In the formalism of canonical quantization, we still have (\ref{eq:cancommu}). Although $a^{(J)}_{n, \vec p_1} (q)$ and $a^{(N)}_{m, \vec p_2} (q)$ depend on $q$, their commutator is independent of $q$ and takes the same form as (\ref{eq:aJaNCom})
%\begin{equation} \label{eq:acommuq}
     %\left[ a^{(J)}_{n, \vec p_1} (q), a^{(N)}_{m, \vec p_2} (q) \right] = (-)^n (2 \pi)^3 \delta_{m+n, 0} \delta^2 (\vec p_1 + \vec p_2) \quad (\forall q \geq 0)\,,
%\end{equation}
%The formula \eqref{eq:acommuq} recovers \eqref{eq:cancommu}, which can be checked by computing the commutators in \eqref{eq:cancommu} through the mode expansion \eqref{eq:JNmodeq}.

%Here the field 
The interacting field $\phi$ also satisfies the Heisenberg equation
\begin{equation}
    \partial_q \phi=-[H_q,\phi]\,,
\end{equation}
where the Hamiltonian for interacting real scalar theory is given by
\begin{equation}
    \begin{split}
        H_q & =H_q^{(0)}+H_q^{(\text{int})} \\
        & =\frac 12 \int \td \psi \td^2 \vec x \left[ - \frac{1}{q} \pi_q^2 - \frac{1}{q} (\p_\psi \phi)^2 + (\vec \p \phi)^2 + m^2 \phi^2 \right] + \int q\td\psi\td^2\vec x V(\phi)\,.
    \end{split}
\end{equation}
Therefore, the interacting field $\phi$ evolves as
\begin{equation}
    \phi (q,\psi, \vec x) = U^{-1} (q, q_0) \phi (q_0, \psi, \vec x) U (q, q_0)
\end{equation}
where the evolution operator $U (q, q_0)$ satisfies the Schr$\Ddot{\text o}$dinger equation
\begin{equation}
    \p_q U (q, q_0) = U (q, q_0) H_q,
\end{equation}
the solution of which is
\begin{equation}
    U (q, q_0) = \mathcal{Q}^{-1} \text{exp} \left[\int^{q}_{q_0} H_{q'} \td q' \right].
\end{equation}
The evolution operator can also be expressed by using the Hamiltonian in the Schr$\Ddot{\text o}$dinger picture. This is similar to the discussion in section \ref{sec:H0U0}, and we will not repeat it here.
%As usual, the reality condition also yields
%\begin{equation}
    %a^{(J) *}_{n, \vec{p}} (q) = (-)^n a^{(J)}_{- n,- \vec{p}} (q) \, , \quad a^{(N) *}_{n, \vec{p}} (q) = (-)^n a^{(N)}_{-n, - \vec{p}} (q) \,.
%\end{equation}
%But now the coefficients $a^{(J)}(q), a^{(N)}(q)$ are no longer independent of the coordinate $q$ due to the interactions $H_q^{(int)}\equiv V(\phi)$. 

Moreover, at the quantum level, $\phi(q\rightarrow0)|0\>$ also needs to be finite, leading to
\begin{equation} %\label{eq:regularint}
    a^{(N)}_{n, \vec p} (q = 0) \ket{0} = 0 \qquad \forall n, \vec p\,,
\end{equation}
or equivalently,
\begin{equation} \label{eq:regularint}
    a^{(N)}_{\theta, \vec p} (q = 0) \ket{0} = 0 \qquad \forall \theta, \vec p\,.
\end{equation}
In general, $a^{(N)}_{\theta, \vec p} (q = \infty) \ket{0} \neq 0$ as $a^{(N)}_{\theta, \vec p} (\infty)$ differs from $a_{\theta, \vec p}^{(N)} (0)$ in a non-trivial way. %we will show very soon in (\ref{eq:diffaNinfaN0}). %Therefore, $a^{(N)}_{n, \vec p} (\infty)$ does not annihilate the vacuum, unless the theory is free. 
%but we do not have (\ref{eq:aJaNCom}) here because
%we can't obtain the coefficients $a^{(J),(N)}_{n,\vec p}(q)$ for a general $q$ through the Klein-Gordon product.
This can be shown by computing the following Klein-Gordon inner product\footnote{Note that the Klein Gordon product $\<\phi_1,\phi_2\>\Big|_q$ is not independent of $q$ for any $\phi_1, \phi_2$ satisfying the equation of motion with non-trivial potential $V(\phi)$ because the dual of the Klein-Gordon current $J^{1 2}$ is no longer closed here, $\td*J^{1 2}\neq0$.} %For example, the Klein product
\begin{equation}
    \< e^{- i p \cdot x}, \phi (x) \>\Big|_q = \sqrt{\frac{\pi}{2}} \sum_{n \in \mathbb Z} i^n e^{i n \theta} J_n (\omega_{\vec p} q) \overset{\longleftrightarrow}{q\p_q} \left[ a^{(J)}_{n, \vec p} (q) J_n (\omega_{\vec p} q) + a^{(N)}_{n, \vec p} (q) N_n (\omega_{\vec p} q) \right].
\end{equation}
%does not lead to the coefficient $a^{(N)}_{\theta,\vec p}(q)$. 
Fortunately, through the asymptotic behavior of the Bessel functions \eqref{eq:asympJNH}, one can find that
\begin{equation} \label{eq:aNintasym}
    a^{(N)}_{\theta, \vec p} (q) \equiv \sqrt{\frac{2}{\pi}} \sum_{n \in \mathbb Z} i^n e^{i n \theta} a^{(N)}_{n, \vec p} (q) = \< e^{- i p x}, \phi (x) \>\Big|_q\,, \quad \text{when} \ q \rightarrow 0, \infty.
\end{equation}
%Similarly, we can have
%\begin{equation}
    %a_{n, \vec p}^{(J)} (q) = -\sqrt{\frac{\pi}{2}}\left\< e^{-i (\vec p \cdot \vec x - n \psi)} N_n (\omega_{\vec p} q), \phi (q, \psi, \vec x) \right\>\Big|_q\,, \quad \text{when} \ q \rightarrow 0, \infty\,,
%\end{equation}
%and
%\begin{equation}
    %a_{n, \vec p}^{(N)} (q) = \sqrt{\frac{\pi}{2}}\left\< e^{-i (\vec p \cdot \vec x - n \psi)} J_n (\omega_{\vec p} q), \phi (q, \psi, \vec x) \right\>\Big|_q\,, \quad \text{when} \ q \rightarrow 0, \infty\,.
%\end{equation}
Now with the Klein-Gordon product at the asymptotic region $(q\rightarrow 0,\infty)$, we can calculate the difference between $a^{(N)}_{\theta, \vec p} (\infty)$ and $a_{\theta, \vec p}^{(N)} (0)$ as
\begin{equation}\label{eq:diffaNinfaN0}
    \begin{split}
        a^{(N)}_{\theta, \vec p} (\infty)-a^{(N)}_{\theta, \vec p} (0)
         =&\<e^{-i p x}, \phi(x) \> \Big|_{q = 0}^{\infty}
         = \int \td \psi \td^2 \vec x \left[ e^{- i p x} \overset{\longleftrightarrow}{q \p_q} \phi (x) \right] \Big|_{q = 0}^{\infty} \\
         %=& \int \td q \td \psi \td^2 \vec x e^{-i p x} \left[ \p_q (q \p_q) - i \omega_{\vec p} \cos \psi + \omega_{\vec p}^2 q \cos^2 \psi \right] \phi (q, \psi, \vec x) \\
         =&\int \td^4 x e^{-i p x} (- \p^2 + m^2) \phi (x) \,.
    \end{split}
\end{equation}
The difference will be important in deriving the LSZ reduction formula.

The scattering amplitude is defined by the inner product between the scattering state living on the boundary of Klein space and the S-vector $|S\rangle\equiv |0\rangle$ (originally mentioned in section \ref{sec:RevKlein}), or more explicitly
\begin{equation}
    \mathcal{A}_n := \langle p_1, \cdots ,p_n|S\rangle = \langle \infty|a_{p_1}(\infty) \cdots a_{p_n}(\infty)|0\rangle,
\end{equation}
where the scattering state defined at the asymptotic region $q \rightarrow \infty$ is given by
\begin{equation}\label{eq:AsymScattStat}
    \begin{aligned}
        \langle p_1, \cdots, p_n|&=\langle \infty|a_{p_1}(\infty) \cdots a_{p_n}(\infty).%\\
        %&=\langle \infty|a_{p_1}(\infty) \cdots a_{p_n}(\infty)U^{-1}(\infty,0)\,,
    \end{aligned}
\end{equation}
 %The explicit expression for operator $a_{p_i}(q=0)$ is not known yet, but its definition at the asymptotic region is clear, 
The explicit expression for the counterpart of $a_p$ in Minkowski space is the Klein-Gordon inner product between the plane wave and the field $\phi (x)$. In Klein space, $a_p$ can be similarly defined as the plane wave generator $\<e^{-i p x}, \phi (x)\>$ in the asymptotic regions $q \rightarrow 0, \infty$,
\begin{equation} \label{eq:aInScattering}
    a_{p}(q) := \< e^{- i p  x}, \phi (x) \> = a^{(N)}_{\theta,\vec{p}}(q), \quad \text{when}~~ q \rightarrow 0, \infty,
\end{equation}
where the expression for $a^{(N)}_{\theta,\vec{p}}(q=0,\infty)$ has been shown in \eqref{eq:aNintasym}. Therefore, the amplitude can be rewritten in terms of the Klein-Gordon inner products $\< e^{-i p x}, \phi (x) \>$
%Based on all discussions in this section, it is now safe to rewrite the amplitude in terms of ``$q$''-order correlation functions by LSZ reduction formula. 
\begin{equation} \label{eq:LSZred}
    \begin{aligned}
        \mathcal{A}_n&=\<\infty|\mathcal{Q}\left(a^{(N)}_{\theta_1,\vec p_1}(\infty)-a^{(N)}_{\theta_1,\vec p_1}(0)\right)\cdots\left(a^{(N)}_{\theta_n,\vec p_n}(\infty)-a^{(N)}_{\theta_n,\vec p_n}(0)\right)|0\>\\
        &=\langle \infty| \mathcal{Q} \< e^{-i p_1 x_1}, \phi (x_1) \> \Big|_{q = 0}^{\infty} \cdots \< e^{-i p_n x_n}, \phi (x_n) \> \Big|_{q = 0}^{\infty}|0\rangle\\
        %&= \left[ \prod^n_{k = 1} \int d^4 x_k e^{-i p_k x_k}(-\p_k^2+m^2) \right] \langle \infty|Q \phi(x_1) \cdots \phi(x_n)|0\rangle\,.
    \end{aligned}
\end{equation}
where we use the regularity condition \eqref{eq:regularint} to add the $a^{(N)}_{\theta_i,\vec p_i}(0) \ket{0}$ term. Note that the $a^{(N)}_{\theta_i,\vec p_i}(\infty)$ generally do not commute with $a^{(N)}_{\theta_i,\vec p_i}(0)$, because $a^{(N)}_{\theta_i,\vec p_i}(\infty)$ are expressed in terms of $a^{(N)}_{\theta_i,\vec p_i}(0), a^{(J)}_{\theta_i,\vec p_i}(0)$. Therefore, the $q$-ordering operator $\mathcal{Q}$ is necessary to move all $a^{(N)}_{\theta_i,\vec p_i}(0)$ attached to $\ket{0}$. Finally, by using the equation (\ref{eq:diffaNinfaN0}), we have
%by using the integral by parts for $\psi, \vec x$, one can derive the following formula for an arbitrary function $F (x)$
\begin{equation}
    \mathcal{A}_n = \left[ \prod^n_{k = 1} \int \td^4 x_k e^{-i p_k x_k}(-\p_k^2+m^2) \right] \langle \infty|\mathcal{Q} \phi(x_1) \cdots \phi(x_n)|0\rangle\,.
\end{equation}
This is in the same form as the LSZ reduction formula in Minkowski space. In particular, in the case of free theory, in which $a^{(N)}_{n, \vec p} (q=0)=a^{(N)}_{n, \vec p} (q=\infty)$, the amplitude will vanish.

\section{Path-integral formalism} \label{sec:PathInt}

In this section, we will study a generally interacting real scalar field in the path-integral formalism. We will start from the path-integral formalism in Euclidean space $(x^0_E, x^1_E, x^2, x^3)$ to define the ket vacuum state $|0\>$, the bra vacuum $\<\infty|$, and the $q$-evolution operator $U(q,q')$, and then turn to Klein space by implementing the Wick rotation $q_E = \sqrt{(x^0_E)^2 + (x^1_E)^2} \rightarrow i q (1 - i \epsilon)$. The canonical quantization results will be reproduced in terms of the path integral.

\begin{figure}[htbp]
    \centering
    \subfigure[The definitions of $\bra{\infty}$ and $\ket{0}$. \label{fig:pathVacqE}]{
    \includegraphics[width=0.45\linewidth]{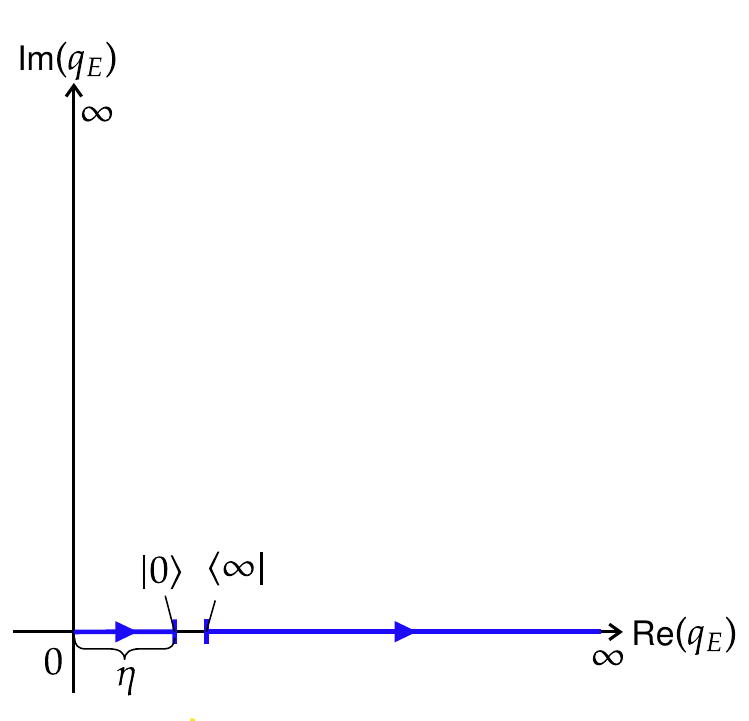}
    }
    \subfigure[The $q$-ordered $n$-point correlation function $\bra{\infty} \phi_n (x_n) \cdots \phi_1 (x_1) \ket{0}$, where $q_1 < \cdots < q_n$. \label{fig:pathCorreqE}]{
    \includegraphics[width=0.45\linewidth]{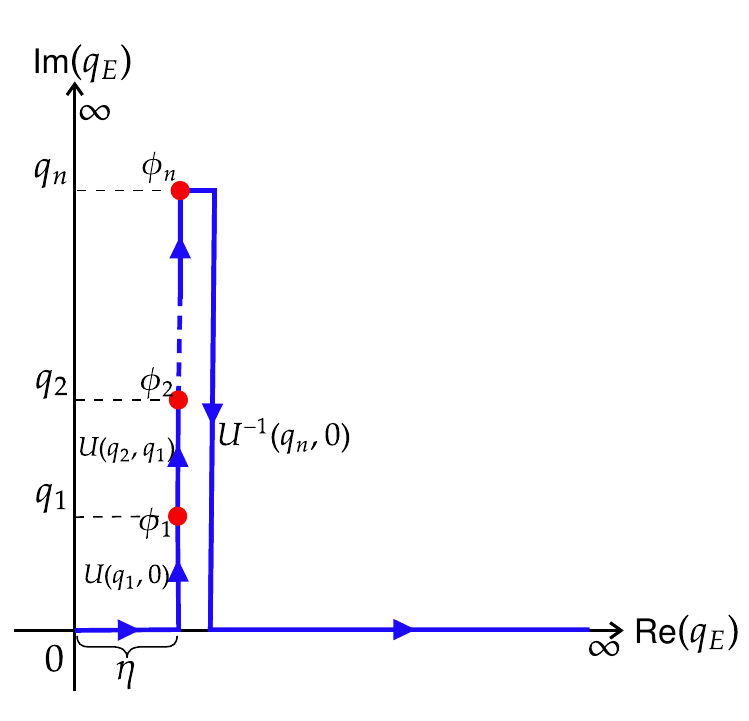}
    }
    \caption{The path integral approach to define vacuum states and correlation functions in the Euclidean signature. The complex plane here is described by the complexified $q_E$.}
    %\label{fig:pathInt}
\end{figure}
\begin{figure}[htbp]
    \centering
    \subfigure[The definitions of $\bra{\infty}$ and $\ket{0}$.]{
    \includegraphics[width=0.45\linewidth]{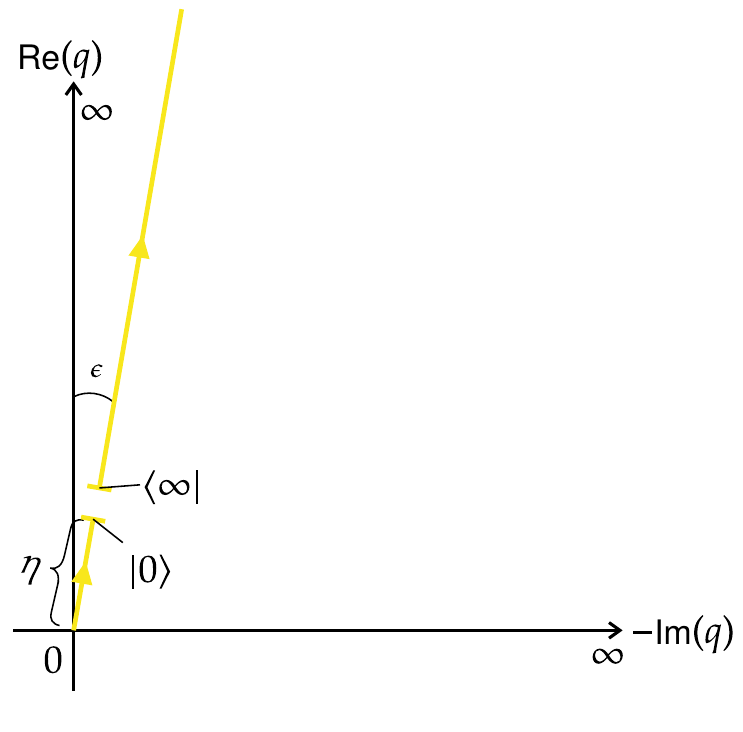}
    }
    \subfigure[The $q$-ordered $n$-point correlation function $\bra{\infty} \phi_n (x_n) \cdots \phi_1 (x_1) \ket{0}$, where $q_1 < \cdots < q_n$.]{
    \includegraphics[width=0.45\linewidth]{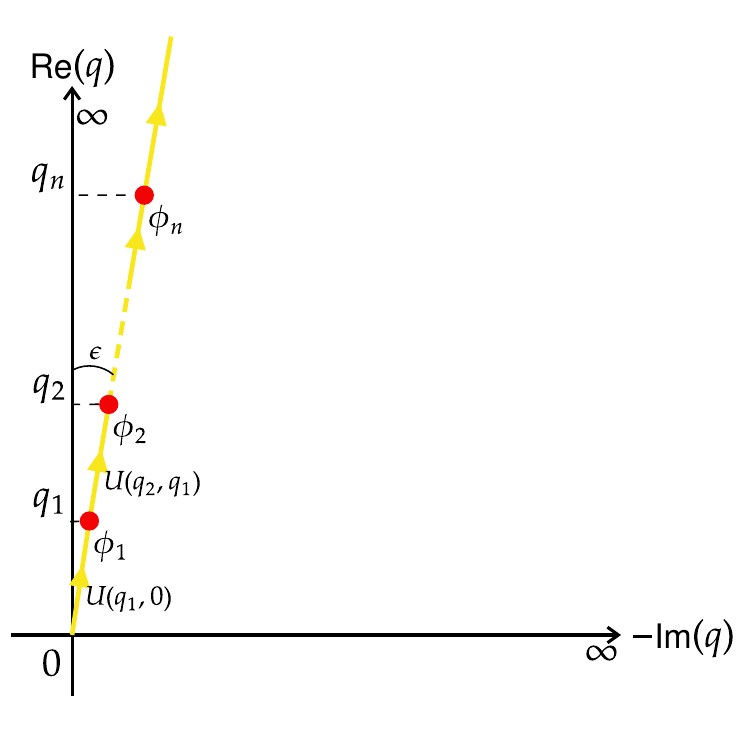}
    }
    \caption{The path integral approach to define vacuum states and correlation functions in the Klein signature. The complex plane here is described by the complexified $q$ with $q= - iq_E$}
    \label{fig:pathInt}
\end{figure}

A general state $|\Psi\>$ in a quantum field theory is a wave functional which maps a field $\tilde{\phi}(\psi,\vec{x})$ to a number $\<\tilde{\phi}|\Psi\>$, where the basis $|\tilde{\phi}\>$ satisfys $\phi(-i\eta,\psi,\vec x)|\tilde{\phi}\>=\tilde{\phi}(\psi,\vec x)|\tilde{\phi}\>$. And then
\begin{equation}
    \begin{aligned}
        &\phi(q-i\eta,\psi,\vec x)|q,\tilde{\phi}\>=U^{-1}(q,0)\phi(-i\eta,\psi,\vec x)U(q,0)U^{-1}(q,0)|\tilde{\phi}\>=U^{-1}(q,0)\tilde{\phi}(\psi,\vec x)|\tilde{\phi}\>\,,\\
        &\<q,\tilde{\phi}|\phi(q-i\eta,\psi,\vec x)=\<\tilde{\phi}|U(q,0)U^{-1}(q,0)\phi(-i\eta,\psi,\vec x)U(q,0)=\<\tilde{\phi}|\tilde{\phi}(\psi,\vec x)U(q,0)\,,
    \end{aligned}
\end{equation}
with $|q,\tilde{\phi}\>\equiv U^{-1}(q,0)|\tilde{\phi}\>$ and $\<q,\tilde{\phi}|=\<\tilde{\phi}|U(q,0)$.
The state $|0\rangle$ and $\bra{\infty}$ are produced by the Euclidean path integral as
\begin{equation}\label{eq:DefVacu}
    \<\tilde{\phi}|0\rangle=\frac{1}{Z_0}\int^{\phi(q_E=\eta,\psi,\vec{x})=\tilde{\phi}(\psi,\vec{x})} D\phi\, e^{\int_{q_E=0}^{\eta} q_E\td q_E \td \psi \td^2\vec{x} \mathcal{L}}\,,
\end{equation}
and
\begin{equation}\label{eq:DefInftyStat}
    \langle \infty|\tilde{\phi}\>=\frac{1}{Z_{\infty}}\int_{\phi(q_E =\eta,\psi,\vec{x})=\tilde{\phi}(\psi,\vec{x})} D\phi\, e^{\int_{q_E = \eta}^{\infty} q_E\td q_E \td \psi \td^2\vec{x} \mathcal{L}}\,,
\end{equation}
where $\eta$ serves as a regulator that keeps $q$ away from the original point, and is ultimately taken to zero in calculating correlation functions. The path-integral definition for the states $\bra{\infty}$ and $\ket{0}$ is illustrated in Fig. \ref{fig:pathVacqE}. The states $\bra{\infty}$ and $\ket{0}$ are located at the boundaries $q_E = \eta$ of the two blue lines. Note that $\<\tilde{\phi}|\phi(q=-i\eta,\psi,\vec{x})|0\>=\tilde{\phi}(\psi,\vec{x})\<\tilde{\phi}|0\>$ is always finite for any function\footnote{Even though sometimes the function $\tilde{\phi}$ can be chosen as a very large value, the divergence will be exponentially suppressed by imposing the boundary condition in the path integral of $\<\tilde{\phi}|0\>$.} $\tilde{\phi}(\psi,\vec{x})$. Then, the finiteness of $\phi(x)|0\>|_{q\rightarrow0}$ is recovered by taking the regulator $\eta$ to zero. Consequently, the definition \eqref{eq:DefVacu} for the vacuum state is consistent with the regularity condition \eqref{eq:regularint}. 

The evolution operator $U(q,q')$ can also be produced by a path integral
\begin{equation}
    \<\tilde{\phi}|U(q,q')|\tilde{\phi}'\>=\<q,\tilde{\phi}|q',\tilde{\phi}'\>=\int_{\phi(q'-i\eta,\psi,\vec{x})=\tilde{\phi}'(\psi,\vec{x})}^{\phi(q-i\eta,\psi,\vec{x})=\tilde{\phi}(\psi,\vec{x})} D\phi\, e^{-\int_{q'-i\eta}^{q-i\eta} q\td q\td\psi\td^2 \vec{x} \mathcal{L}}\,.
\end{equation}
The path integral for a series of evolution operators $U(q_1,0), U(q_2,q_1),\cdots, U^{-1}(q_n,0)$ can be depicted in Fig. \ref{fig:pathCorreqE}, wherein the evolution operators are represented by some vertical blue lines with boundary at %$q = 0-i\eta,q_1-i\eta,q_2-i\eta,\cdots,q_n-i\eta$
$q_E = \eta, \eta + i q_1, \eta + i q_2,\cdots, \eta + i q_n$.
Then, the $q$-ordered correlation function $\<\infty|\mathcal{Q} \phi(x_1) \cdots \phi(x_n)|0\>$ can be represented by a row of red points connected by blue lines, where the red points denote the fields.
Moreover, the path-integral formalism can be equivalently reformulated in the Klein signature by Wick rotation, which is illustrated by the yellow lines in Fig. \ref{fig:pathInt}. 
As a result, the path-integral definition of $n$-point correlation function is 
\begin{equation}\label{eq:CorrFuncPathInt}
    \<\infty| \mathcal{Q} \phi(x_1) \cdots \phi(x_n)|0\>=\frac{1}{Z}\int D \phi\, \phi(x_1) \cdots \phi(x_n) e^{-\int_{q=0}^{\infty(1-i\epsilon)} q\td q \td\psi\td^2\vec{x} \mathcal{L}}\,,
\end{equation}
where $Z=Z_0 Z_{\infty}$. With the infinitesimal constant $\epsilon$, the right-hand side will be automatically $q$-ordered. It is also important to stress that, similar to Feynman's $i\epsilon$-prescription in the Minkowski signature, the $i\epsilon$-prescription is also needed for the path-integral formalism in the Klein signature, as the path integral without $\epsilon$ is not well defined.
In particular, the vacuum expectation is 
\begin{equation}\label{eq:VacuCorrFunc}
    \<\infty|0\>=\frac{1}{Z}\int D \phi\, e^{-\int_{q=0}^{\infty(1-i\epsilon)} q\td q \td\psi\td^2\vec{x} \mathcal{L}}\,.
\end{equation}
Without generality, we can choose the normalization $\<\infty|0\>=1$, with the normalization constant $Z$ satisfying,
\begin{equation}
    Z=\int D \phi\, e^{-\int_{q=0}^{\infty(1-i\epsilon)} q\td q \td\psi\td^2\vec{x} \mathcal{L}}\,.
\end{equation}

These results exactly match the results from direct analytic continuation as shown in Eq. \eqref{eq:AnaContiPathInt}. As an example, 
we can consider the two-point function $\<\infty|\mathcal{Q}\phi(x)\phi(x')|0\>$ of free field. Instead of  calculating the path integral directly
\begin{equation}
    \<\infty| \mathcal{Q} \phi(x) \phi(x')|0\>=\frac{1}{Z}\int D \phi\, \phi(x) \phi(x') e^{-\int \td^4 x \mathcal{L}_0},,
\end{equation}
we compute the vacuum partition functional with a source $J(x)$,
\begin{equation}
    Z[J]=\int D\phi e^{-\int \td^4 x (\mathcal{L}_0+J\phi)}\equiv\int D\phi e^{-S[J]}\,.
\end{equation}
Defining $\tilde{\phi}(p)=\int \td^4 x e^{i p x}\phi(x)$ and $\tilde{J}(p)=\int \td^4 x e^{i p x}J(x)$, we obtain
\begin{equation}
    \begin{aligned}
        S[J]=&\frac{1}{2}\int\frac{\td^4 p}{(2\pi)^4}[-\tilde{\phi}(p)(p^2+m^2)\tilde{\phi}(-p)+\tilde{J}(p)\tilde{\phi}(-p)+\tilde{J}(-p)\tilde{\phi}(p)]\\
        =&\frac{1}{2}\int\frac{\td^4 p}{(2\pi)^4}\left[\frac{\tilde{J}(p)\tilde{J}(-p)}{p^2+m^2}-\tilde{\chi}(p)(p^2+m^2)\tilde{\chi}(-p)\right]\,,
    \end{aligned}
\end{equation}
with $\tilde{\chi}(p)=\tilde{\phi}(p)-\frac{\tilde{J}(p)}{p^2+m^2}$. Then we have
\begin{equation}
    \begin{aligned}
        Z[J]=Z &\exp\left[-\frac{1}{2}\int\frac{\td^4 p}{(2\pi)^4} \frac{\tilde{J}(p)\tilde{J}(-p)}{p^2+m^2-i\epsilon} \right]\\
        =Z &\exp\left[\frac{1}{2}\int\td^4 x\td^4 x' J(x)\Delta(x-x')J(x') \right]
    \end{aligned}
\end{equation}
with  
\begin{equation}
    \Delta(x-x')=-\int\frac{\td^4 p}{(2\pi)^4}\frac{e^{i p (x-x')}}{p^2+m^2-i\epsilon}
\end{equation}
being the two-point function of the free field.

\section{Conclusions and discussions}\label{sec:ConDiss}

In this work, we investigated the canonical quantization and the path-integral quantization of a scalar field theory in Klein space. In canonical quantization, we selected the ``length of time" $q$ as the evolution parameter such that the field can be expanded in terms of the Bessel functions and the Neumann functions. Though the Neumann functions are divergent at the original point, they are crucial in canonical quantization. We imposed the regularity condition to constrain the Neumann vacuum state $|0\>$. On the other hand, we also define the Hankel vacuum state $\<\infty|$ which is associated with the asymptotic region $q\rightarrow\infty$, where the mode expansions behave as $e^{\pm i \omega_{\vec p}q}$. By using these constructions, we explicitly calculated the free two-point function, which is consistent with the covariant one. For general $q$-ordered correlation functions with small interactions, they can be calculated perturbatively. The details can be found in appendix \ref{ap:PertExpand}. And then we deduced the LSZ reduction formula in the Heisenberg picture. Furthermore, we reconstructed the states and the correlation functions in the framework of path-integral quantization.

%The modes related to the former are in fact the modes of plane waves, while the modes related to the latter are divergent at the original point, which is beyond any modes we met in a usual QFT textbook. However, the Neumann function modes we introduced turned out to be crucial in the discussion of the canonical quantization and the LSZ reduction formula for a QFT in Klein space. Finally, we can define every state and operator in the formalism of canonical quantization and in the formalism of path integral, and thereafter calculate a general $q$-ordered correlation function. And we managed to prove the LSZ reduction formula, which translates the amplitudes to the $q$-ordered correlation functions. It is interesting that our novel construction leads to results exactly the same as the results from direct analytical continuation.}

One subtle point is that the order in $q$ in our formalism depends on the choice of the original point. A different choice of the original point could lead to a different ordering of operators in the correlation functions. In other words, the $q$-ordering operator $\mathcal{Q}$ does not commute with spacetime translations. This is similar to the radial quantization of CFT, where the radial-ordering operator $R$ in the definition of correlation functions does not commute with the spacetime translations. In contrast, the time-ordering operator $T$ in Minkowski spacetime does commute with the spacetime translations (but does not commute with the Lorentz boosts generally). 

The novel modes associated with the Neumann functions need further study. They may play an essential role in flat holography. One unanswered question in flat holography concerns the extrapolation dictionary between the operators in boundary field theory and the bulk fields. It would be interesting to investigate the role of these novel modes in the bulk reconstruction in flat holography.

Apart from the scalar field, a far more interesting topic is the amplitudes of gauge fields in Klein space, where the nontrivial on-shell three-point amplitude is the basics of higher-point amplitudes as shown in the BCFW recursion relation \cite{Britto:2005fq}. We believe the construction here can be generalized to gauge theories with some effort because the gauge symmetry is perturbatively irrelevant to the Lorentz spacetime symmetry we treated here. The semi-classical quantization around some non-perturbative objects, like magnetic monopoles, may exhibit more interesting behavior. It would also be interesting to generalize our study to spinors and tensors.

%may cause concern from the flat holography perspective. A naive statement in the flat holography from the Carrollian approach is that the operators in the boundary Carrollian conformal field theory one-to-one correspond to the bulk fields approaching the null infinity, which is often called the extrapolation dictionary. In the story of flat holography in Minkowski spacetime, the boundary field operators correspond to the annihilation/creation operators accompanied by the modes of the plane waves. Thus, the role that the brand new modes in flat holography of Klein space play needs to be figured out. We hope to answer that in the near future.）

As a final remark, most of our discussions apply to a more general spacetime $\mathbb{R}^{n,m} (n,m\geq2)$, which also has only one asymptotic boundary. Although the solution basis of the Klein-Gordon equation would be different, the basic physical picture would remain the same, we will discuss this in \cite{Chen:2025eeh}. We can still define ``the length of time" and introduce additional modes beyond plane waves in the associated coordinate system. The canonical quantization, the path integral quantization, and the LSZ reduction would also be realizable.

%\mxc{The arguments for finite $q$:...}
%\section*{Acknowledgments}
%\cb{We thank Yu-fan Zheng very much for the valuable discussions. }

\acknowledgments
We would like to thank Yu-fan Zheng for inspiring discussions and thank Yu-ting Wen, Jie Xu, and Zhi-jun Yin for the valuable suggestions on the manuscript. This research is supported in part by NSFC Grant  No. 11735001, 12275004, 12588101. 

\appendix

\section{Amplitudes and correlation functions from analytical continuation}\label{ap:analyCon}

In this section, we derive the ``time"-ordered correlation functions and LSZ reduction formula by doing direct analytical continuation from the Euclidean and Minkowski spacetime.

First, consider a QFT in Euclidean space
\begin{equation}
     d s^2  =  (d x^0)^2 + (d x^1)^2 + (d x^2)^2 + (d x^3)^2\,,
\end{equation}
we can tell any correlation function by the path integral
\begin{equation}
    Z [J]  =  \int D \phi e^{\int d^4 x (L + J \phi)}\,,
\end{equation}
where $\int d^4 x L$ is definitely negative to make sure the path integral is well-defined.

It can be analytically continued to Minkowski space
\begin{equation}
    d s^2  =  - (d x^0)^2 + (d x^1)^2 + (d x^2)^2 + (d x^3)^2
\end{equation}
by Wick rotation
\begin{equation}
    x^0 \rightarrow i x^0 (1 - i
\epsilon), p_0 \rightarrow - i p_0 (1 + i \epsilon)\,.
\end{equation}
Then the Feynman integral is given by
\begin{equation}
  Z [J]  =  \int D \phi e^{i \int d^4 x (L + J \phi)} \equiv \int D \phi
  e^{i S}\,.
\end{equation}
Take a massive scalar, for example, its free two-point function is given by
\begin{equation}
    \Delta (x - x')  \equiv  -i\int \frac{d^4 p}{(2 \pi)^4}  \frac{e^{i p (x -
  x')}}{p^2 + m^2 - i \epsilon}
\end{equation}
where the $\epsilon$-prescription comes from $\frac{e^{i k_E  (x_E -
x_E')}}{k_E^2 + m^2}$ by using $x^0 \rightarrow i x^0 (1 - i \epsilon), p_0
\rightarrow - i p_0 (1 + i \epsilon)$.
Then the time-ordered correlations are
\begin{equation}
    \langle 0| T \phi (x_1) \ldots \phi (x'_1) \ldots|0 \rangle 
  =  \frac{\delta}{i \delta J (x_1)} \ldots \frac{\delta}{i \delta J (x'_1)}
  \ldots Z [J] |_{J = 0} \,,
\end{equation}
and the LSZ formula gives the amplitude
\begin{equation}
    \begin{aligned}
        \langle f | i \rangle
        =&  \langle p_{1'} p_{2'} \ldots p_{n'} | S | p_1 p_2 \ldots
        p_n \rangle \\
        = & i^{n + n'} \int d^4 x_1  e^{i p_1 x_2} (- \partial_1^2 + m^2)
        \ldots\\
        &  \times \int d^4 {x'_1}  e^{- i p'_1 x'_2} (- \partial_{1'}^2 + m^2)
        \ldots\\
        &   \times \langle  \phi (x_1) \ldots \phi (x'_1) \ldots \rangle\,.
    \end{aligned}
\end{equation}
Note that we have assumed $p^0 > 0$ for every in-going or out-going particle,
Thus, we can distinguish the in-going or out-going particles by the signature
in $e^{\pm i p x}$, ``$+$" for in-going and ``$-$" for out-going.

We can also analytically continue the QFT from Minkowski space to Klein space
\begin{equation}
    d s^2  =  - (d x^0)^2 - (d x^1)^2 + (d x^2)^2 + (d x^3)^2
\end{equation}
by $x^1 \rightarrow i x^1 (1 - i
\epsilon), p_1 \rightarrow - i p_1 (1 + i \epsilon)$.
The Feynman integral is given by
\begin{equation}\label{eq:AnaContiPathInt}
    Z [J]  =  \int D \phi e^{- \int d^4 x (L + J \phi)} \equiv \int D \phi
  e^{- S}\,.\\
\end{equation}
The two-point function is similar to that in Minkowski space
\begin{equation}
    \Delta (x - x')  \equiv  - \int \frac{d^4 p}{(2 \pi)^4}  \frac{e^{i p (x -
  x')}}{p^2 + m^2 - i \epsilon}\,.
\end{equation}
Then the ``time"-ordered, which turn out to be $q$-ordered in our formalism shown in section \ref{sec:CanonicalQ}, correlation functions are\footnote{The left vector state $\langle \infty|$ is a quite different state from the vacuum state $|0\>$ because we choose to canonical quantize the QFT on a hypersurface depending on $q$ but $\p_q$ is not a (conformal) Killing vector. They are defined explicitly in the section \ref{sec:PathInt}.}
\begin{equation}
    \langle \infty|\mathcal{Q} \phi (x_1) \ldots \phi (x_n) |0 \rangle  
    =  \frac{- \delta}{\delta J (x_1)} \ldots \frac{- \delta}{\delta J (x_n)} Z
    [J] |_{J = 0} .
\end{equation}
We would expect that there is a similar LSZ reduction formula giving
\begin{equation}
    \begin{aligned}
        \mathcal{A} 
   = & \langle p_1 p_2 \ldots p_n  | S  \rangle
  \\
   = & \int d^4 x_1  e^{-i p_1 x_1} (- \partial_1^2 + m^2) \ldots\\
    & \times \langle  \phi (x_1) \ldots \phi (x_n) \rangle\,.
    \end{aligned}
\end{equation}
Here we cannot distinguish the incoming or outgoing particles because the mass-shell condition $p^2=0$ in Klein signature has only one connected component, which reflects the fact that there is only one asymptotic boundary for Klein space.

\section{Self-consistency check of the propagator} \label{ap:propagator}

In this section, we verify the equivalence between the two approaches in deriving the two-point propagator, as presented in equations \eqref{eq:Wickpropagator} and \eqref{Greenfucntion} respectively. Since the expression (\ref{Greenfucntion}) is manifestly invariant under Poincar\'e transformations, we can choose $x' = (0, 0, \vec 0)$ such that
\begin{equation}
    \begin{aligned}
        \frac{\Delta (x)}{\bra{\infty}0\>}&=-\int \frac{\td^4 p}{(2\pi)^4}\frac{e^{i p x}}{p^2+m^2-i\epsilon}\\ 
        & = -\int \frac{\omega \td\omega\td\theta\td^2 \vec p}{(2\pi)^4}\frac{e^{-i\omega q \sin\theta+i \vec p\cdot\vec x}}{-\omega^2+\omega_{\vec p}^2-i\epsilon}\\
        & =-\left(\int_0^\pi \td\theta+\int_{-\pi}^0 \td\theta\right)\int\frac{\omega\td\omega\td^2 \vec p}{(2\pi)^4}\frac{e^{-i\omega q \sin\theta+i \vec p\cdot\vec x}}{-\omega^2+\omega_{\vec p}^2-i\epsilon}.
        %=&2\pi i \int_0^\pi \td\theta\int\frac{\omega_{\vec p}\td^2 \vec p}{(2\pi)^4}\frac{e^{-i\omega_{\vec p} q \sin\theta+i \vec p\cdot\vec x}}{-2\omega_{\vec p}}+\int_0^\pi \td\theta \int_{-i\infty}^{0}\frac{\omega\td\omega\td^2 \vec p}{(2\pi)^4}\frac{e^{-i\omega q \sin\theta+i \vec p\cdot\vec x}}{-\omega^2+\omega_{\vec p}^2}\\&-\int_{-\pi}^0 \td\theta\int_{0}^{i\infty}\frac{\omega\td\omega\td^2 \vec p}{(2\pi)^4}\frac{e^{-i\omega q \sin\theta+i \vec p\cdot\vec x}}{-\omega^2+\omega_{\vec p}^2}\\
        %& = \int\frac{\td^2\vec p}{(2\pi)^3}e^{i \vec p \cdot\vec x} \mathcal{I}_{\vec p} %\left(-\frac{i}{2} \int_0^\pi\td\theta e^{-i\omega_{\vec p}q\sin\theta}+2 \int_{0}^{\pi} \td\theta\int_{0}^{\infty}\frac{\omega\td\omega}{2\pi}\frac{e^{-\omega q \sin\theta}}{\omega^2+\omega_{\vec p}^2}  \right)\\
        %=&\int\frac{\td^2\vec p}{(2\pi)^3}e^{i \vec p \cdot\vec x}\bigg(-\frac{i\pi}{2}\left(BesselJ[0,\omega_{\vec p} q]-i StruveH[0,\omega_{\vec p} q]\right)\\
        %&+\frac{1}{2}\left(BesselJ[0,\omega_{\vec p} q] 2log(2/\omega_{\vec p} q)+\pi StruveH[0,\omega_{\vec p} q]-2 Hypergeometric0F1Regularized^{(1,0)}[1,-(\omega_{\vec p} q)^2/4]  \right)   \bigg)
    \end{aligned}
\end{equation}
To proceed with the above integral represented by a red line interval in Fig. \ref{fig:diag-integral-propagator}, we rotate the two parts $\theta>0$ and $\theta<0$ into different half imaginary coordinate axes represented by the blue line and the purple line, respectively. When rotating to the blue line, we cross a pole at $\omega=\omega_{\vec p}-i\epsilon$ and obtain a contribution by using the residue theorem.
\begin{figure}
    \centering
    \includegraphics[width=0.4\linewidth]{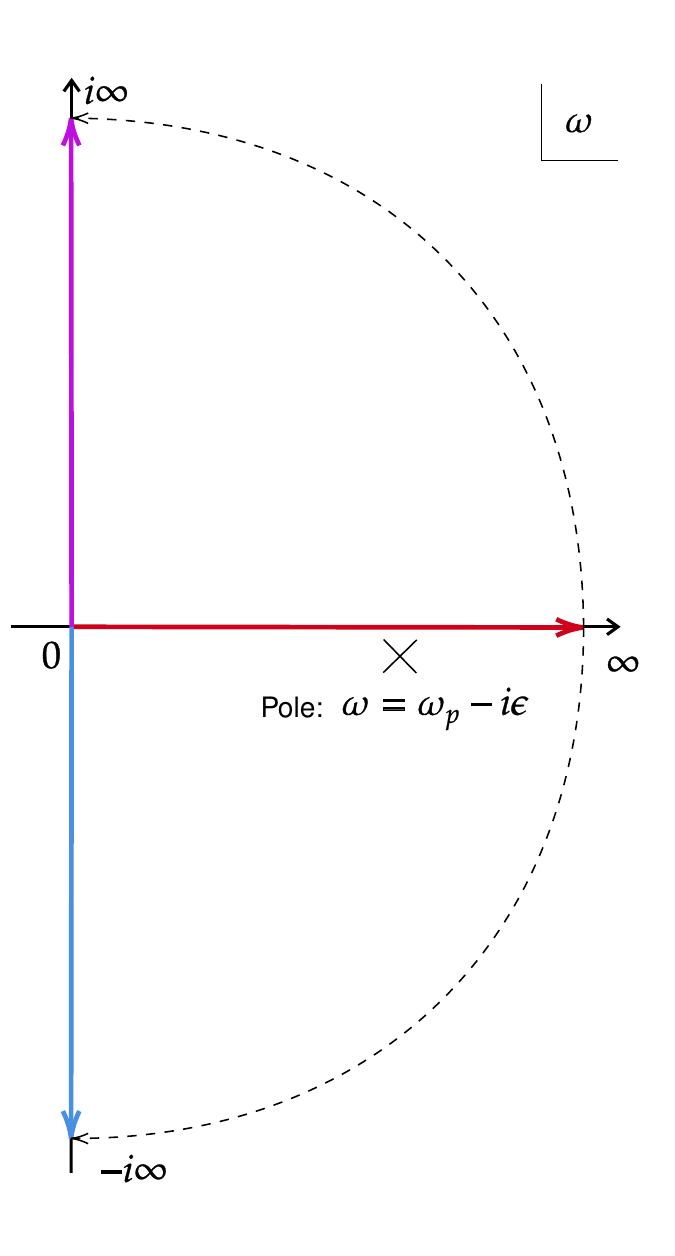}
    \caption{The integral contour for the complexified $\omega$.}
    \label{fig:diag-integral-propagator}
\end{figure}
Then the final result is
\begin{equation}
    \frac{\Delta (x)}{\bra{\infty}0\>} = \int\frac{\td^2\vec p}{(2\pi)^3}e^{i \vec p \cdot\vec x} \mathcal{I}_{\vec p}
\end{equation}
where
\begin{equation}
    \mathcal{I}_{\vec p} = -\frac{i}{2} \int_0^\pi\td\theta e^{-i\omega_{\vec p}q\sin\theta}+2 \int_{0}^{\pi} \td\theta\int_{0}^{\infty}\frac{\omega\td\omega}{2\pi}\frac{e^{-\omega q \sin\theta}}{\omega^2+\omega_{\vec p}^2}.
\end{equation}
These two integrals can be calculated by using Mathematica
\begin{equation}
    \begin{split}
        \mathcal{I}_{\vec p} = & -\frac{i\pi}{2}\left( J_0 (\omega_{\vec p} q) - i  H_0 ( \omega_{\vec p} q ) \right) + J_0 (\omega_{\vec p} q) \ln \frac{2}{\omega_{\vec p} q} \\
        & + \frac{\pi}{2} H_0 ( \omega_{\vec p} q ) - \lim_{\alpha \rightarrow 0} \p_\alpha \left[ \left(\frac{2}{\omega_{\vec p} q} \right)^\alpha J_\alpha (\omega_{\vec p} q) \right],
    \end{split}
\end{equation}
where $H_0 (x)$ denotes the Struve function and can be given in terms of a power series
\begin{equation}
    H_0 (x) = \sum_{m = 0}^\infty \frac{(-)^m}{\left[ \Gamma (m + \frac32) \right]^2} \left( \frac{x}{2} \right)^{2m + 1}.
\end{equation}
It can be verified that 
\begin{equation} \label{eq:In2}
    \mathcal{I}_{\vec p} = -\frac{i\pi}{2}H_0^{(2)}(\omega_{\vec p} q),
\end{equation}
where $H_0^{(2)}$ is the second kind Hankel function.
Therefore, \eqref{Greenfucntion} can be rewritten as
\begin{equation}
    \Delta (x - x') = -\frac{i\pi}{2} \int\frac{\td^2\vec p}{(2\pi)^3}e^{i \vec p \cdot\vec x} H_0^{(2)}(\omega_{\vec p} \Delta q) \bra{\infty}0\>.
\end{equation}
This matches the result we derived from the Wick contraction \eqref{eq:Wickpropagator}.

\section{Perturbative expansion and Wick theorem}\label{ap:PertExpand}

In this appendix, we will show how to perturbatively calculate an interactive real scalar field theory in Klein space within the canonical quantization formalism.

In the interaction picture, the scalar field behaves the same as in the free case,
\begin{equation}
    \phi_{I}(q)=U^{-1}_{(0)}(q,q_0)\phi_I(q_0)U_{(0)}(q,q_0)\,,
\end{equation}
which is related to the field in the Heisenberg picture as
\begin{equation}
    \phi(q)=U^{-1}_I(q,q_0)\phi_I(q) U_I(q,q_0)\,,
\end{equation}
where the evolution operator in the interaction picture is defined to be
\begin{equation}
    U_I(q,q_0)=U^{-1}_{(0)}(q,q_0)U(q,q_0)\,.
\end{equation}
From the definition of $U_I(q,q_0)$, we have
\begin{equation}
    \partial_q U_I(q,q_0)=H_I(q) U_I(q,q_0)\,,
\end{equation}
the solution to which is
\begin{equation}
    U_I(q,q_0)=\mathcal{Q}\exp\left[\int^{q}_{q_0}\td q H_I(q)\right]\,,
\end{equation}
where the Hamiltonian in the interaction picture $H_I(q)$ is
\begin{equation}
    H_I(q)=U^{-1}_{(0)}(q,q_0)\left(H_{q}^{(S)}-H_{q}^{(0)(S)}\right)U_{(0)}(q,q_0)=\int q \td\psi
    \td^2 \vec x V(\phi_I(q))\,.
\end{equation}

Due to the interaction $V(\phi)$, the Hankel vacuum state should be adjusted to 
\begin{equation}
    \<\infty|\propto\<\infty,\text{free}|U_I(\infty,q_0)\,,
\end{equation} 
while the Neumann vacuum state becomes 
\begin{equation}
    |0\>\propto U_I(q_0,0)|0,\text{free}\>\,,
\end{equation}
where the free states $|0,\text{free}\>$, $\<\infty,\text{free}|$ are defined as (\ref{eq:regular}) and (\ref{eq:InftyStateAnnihilation}) respectively in the free theory case. And it is convenient to choose $q_0=0$. Thus, a general correlation function $\<\infty|\phi(q_n)\cdots\phi(q_1)|0\>$ with $q_n>\cdots>q_1$ can be calculated as
\begin{equation}\nonumber
    \begin{aligned}
        &\<\infty|\phi(q_n)\cdots\phi(q_1)|0\>\\
        \propto&\<\infty,\text{free}|U_I(\infty,0)U^{-1}_I(q_n,0)\phi_I(q_n)U_I(q_n,0)\cdots U^{-1}_I(0,q_1)\phi_I(q_1)U_I(q_1,0)|0,\text{free}\>\\
        =&\<\infty,\text{free}|U_I(\infty,q_n)\phi_I(q_n)U_I(q_n,q_{n-1})\cdots U_I(q_2,q_1)\phi_I(q_1)U_I(q_1,0)|0,\text{free}\>\,,
    \end{aligned}
\end{equation}
which can be expressed in a more compact way as
\begin{equation}\nonumber
    \<\infty|\mathcal{Q}\phi(q_1)\cdots\phi(q_n)|0\>\propto \<\infty,\text{free}|\mathcal{Q}\phi_I(q_1)\cdots\phi_I(q_n)\exp\left[\int^{\infty}_{0}\td q H_I(q)\right]|0,\text{free}\>\,.
\end{equation}
If we choose the normalization $\<\infty|0\>=1$, we have
\begin{equation}\label{eq:PertCorrFuncs}
    \<\infty|\mathcal{Q}\phi(q_1)\cdots\phi(q_n)|0\>=\frac{\<\infty,\text{free}|\mathcal{Q}\phi_I(q_1)\cdots\phi_I(q_n)\exp\left[\int^{\infty}_{0}\td q H_I(q)\right]|0,\text{free}\>}{\<\infty,\text{free}|\mathcal{Q}\exp\left[\int^{\infty}_{0}\td q H_I(q)\right]|0,\text{free}\>}\,.
\end{equation}

Next, we will demonstrate how the Wick theorem still works, as it slightly differs from the process in Minkowski spacetime. Since the free states $|0,\text{free}\>$, $\<\infty,\text{free}|$ are annihilated by $a^{(N)}$ and $a^{(H^{(1)})}$. We need to expand the field $\phi_I$ as
\begin{equation}
    \phi_I(q,\psi,\vec x)=\phi_I^{+}(q,\psi,\vec x)+\phi_I^{-}(q,\psi,\vec x)\,,
\end{equation}
with
\begin{equation}
    \phi_I^{+}(q, \psi, \vec x)  = \sqrt{\frac{\pi}{2}} \sum_{n \in \mathbb Z} \int \frac{\td^2 \vec p}{(2 \pi)^3} \left[ i a^{(N)}_{n, \vec p} H^{(2)}_n ( \omega_{\vec p} q ) \right] e^{i (\vec p \cdot \vec x - n \psi)}
\end{equation}
and
\begin{equation}
    \phi_I^{-}(q, \psi, \vec x)  = \sqrt{\frac{\pi}{2}} \sum_{n \in \mathbb Z} \int \frac{\td^2 \vec p}{(2 \pi)^3} \left[ 2 a^{(H^{(1)})}_{n, \vec p} J_n ( \omega_{\vec p} q ) \right] e^{i (\vec p \cdot \vec x - n \psi)}\,.
\end{equation}
Obviously, we have $\<\infty,\text{free}|\phi^-_I=0$ and $\phi^+_I|0,\text{free}\>=0$. Thus, we define the normal ordering $\mathcal{N}$ as all the $\phi^+_I$'s being to the right of all the $\phi^-_I$'s and define the Wick contraction $\contraction{}{\phi_I}{(q)}{\phi_I}\phi_I(q)\phi_I(q')$ as
\begin{equation}
    \contraction{}{\phi_I}{(q)}{\phi_I}\phi_I(q)\phi_I(q') = 
    \begin{cases}
        & [\phi_I^+(q),\phi_I^-(q')]\,,\quad\text{if}\,q>q'\,,\\
        & [\phi_I^+(q'),\phi_I^-(q)]\,,\quad\text{if}\,q'>q\,.
    \end{cases}
\end{equation}
%\begin{equation}
    %\begin{aligned}
        %[\phi_I(q)\phi_I(q')]&=[\phi_I^+(q),\phi_I^-(q')]\,,\quad\text{if}\,q>q'\,,\\
        %[\phi_I(q)\phi_I(q')]&=[\phi_I^+(q'),\phi_I^-(q)]\,,\quad\text{if}\,q'>q\,.
    %\end{aligned}
%\end{equation}
And this Wick contraction of two fields is just the Feynman propagator $\Delta(x-x')$, which we explicitly calculated in the appendix \ref{ap:propagator}.

Finally,  we can perturbatively expand the exponential $\exp\left[\int^{\infty}_{0}\td q H_I(q)\right]$ in the right-hand side of Eq.(\ref{eq:PertCorrFuncs}) and use the Wick theorem that
\begin{equation}
    \mathcal{Q}\left\{\phi_I(q_1)\phi_I(q_2)\cdots\phi_I(q_n)\right\}=\mathcal{N}\left\{\phi_I(q_1)\phi_I(q_2)\cdots\phi_I(q_n)+\text{all possible contractions}\right\}
\end{equation}
to obtain the final results, where all terms in the expansions can be labeled by Feynman diagrams. 

\section{Causal structure}\label{ap:CausalStr}
In this appendix, we want to demonstrate causal structures of a QFT in Euclidean, Minkowski, and Klein spaces.
We use a very simple model, a conformal field theory (CFT).

The Euclidean correlator reads
\begin{equation}\label{eq:ConfEuc}
    \langle O (x) O (0) \rangle_E =  \frac{1}{(x^2)^{\Delta}}
    =  \frac{1}{[(x^0)^2 + (x^1)^2 + (\vec{x})^2]^{\Delta}}\,,
\end{equation}
with $\vec x=(x^2, x^3)$. It can be analytically continued to Minkwoski space by $x^0 \rightarrow i x^0 (1 \mp i
\epsilon)$.
If $- \sqrt{(x^1)^2 + (\vec{x})^2} < x^0 < \sqrt{(x^1)^2 + (\vec{x})^2}$ , starting from (\ref{eq:ConfEuc}), the
analytic continuation won't pass any branch point, and we have
\begin{equation}\label{eq:ConfMink0}
  \langle O (x) O (0) \rangle_M  =  \frac{1}{[- (x^0)^2 + (x^1)^2 +
  (\vec{x})^2]^{\Delta}}\,.
\end{equation}
If $x^0 > \sqrt{(x^1)^2 + (\vec{x})^2}$ or $x^0 < - \sqrt{(x^1)^2 +
(\vec{x})^2}$, starting from (\ref{eq:ConfEuc}), different ways to avoid the branch point give different
correlators:
\begin{equation}\label{eq:ConfMink1}
    \begin{aligned}
        \langle O (x) O (0) \rangle_E\xrightarrow{x^0 \rightarrow i x^0 (1 - i
        \epsilon)}\langle O (x) O (0) \rangle_M^{(1)} & =  \frac{1}{e^{i \pi \Delta} [(x^0)^2 -
        (x^1)^2 - (\vec{x})^2]^{\Delta}}\,,\\
        \langle O (x) O (0) \rangle_E\xrightarrow{x^0 \rightarrow i x^0 (1 + i
        \epsilon)}\langle O (x) O (0) \rangle_M^{(2)} & =  \frac{1}{e^{- i \pi \Delta} [(x^0)^2 -
        (x^1)^2 - (\vec{x})^2]^{\Delta}}\,,
    \end{aligned}
\end{equation}
where the former one gives the time-ordered correlator and the latter one gives the anti-time-ordered correlator.

Now consider analytic continuation from Minkowski spacetime to Klein space by $x^1 \rightarrow i x^1 (1 \mp i\epsilon)$.
If $x^0 > | \vec{x} |$ or $x^0 < - | \vec{x} |$, starting from (\ref{eq:ConfMink1}), the analytic continuation won't pass any branch point, we have
\begin{equation}
    \begin{aligned}
        \langle O (x) O (0) \rangle_M^{(1)}\xrightarrow{x^0 \rightarrow i x^0 (1 \mp i
        \epsilon)}\langle O (x) O (0) \rangle_K^{(1)}  = & \frac{1}{e^{i \pi \Delta} [(x^0)^2 +
        (x^1)^2 - (\vec{x})^2]^{\Delta}}\,,\\
        \langle O (x) O (0) \rangle_M^{(2)}\xrightarrow{x^0 \rightarrow i x^0 (1 \mp i
        \epsilon)}\langle O (x) O (0) \rangle_K^{(2)}  = & \frac{1}{e^{- i \pi \Delta} [(x^0)^2 +
        (x^1)^2 - (\vec{x})^2]^{\Delta}}\,.
    \end{aligned}
\end{equation}
If $- | \vec{x} | < x^0 < | \vec{x} |$, and if $x^1 > \sqrt{- (x^0)^2 +
(\vec{x})^2}$ or $x^1 < - \sqrt{- (x^0)^2 + (\vec{x})^2}$, starting from (\ref{eq:ConfMink0}), different ways to avoid the branch point give different correlators: 
\begin{equation}
    \begin{aligned}
        \langle O (x) O (0) \rangle_M\xrightarrow{x^0 \rightarrow i x^0 (1 - i
        \epsilon)}\langle O (x) O (0) \rangle_K^{(1)}  = & \frac{1}{e^{i \pi \Delta} [(x^0)^2 +
        (x^1)^2 - (\vec{x})^2]^{\Delta}}\,,\\
         \langle O (x) O (0) \rangle_M\xrightarrow{x^0 \rightarrow i x^0 (1 + i
        \epsilon)}\langle O (x) O (0) \rangle_K^{(2)}  = & \frac{1}{e^{- i \pi \Delta} [(x^0)^2 +
        (x^1)^2 - (\vec{x})^2]^{\Delta}}\,.
    \end{aligned}
\end{equation}
If $- | \vec{x} | < x^0 < | \vec{x} |$, and if $-\sqrt{- (x^0)^2 +
(\vec{x})^2}<x^1<\sqrt{- (x^0)^2 +
(\vec{x})^2}$, starting from (\ref{eq:ConfMink0}), it won't pass any branch point, we have
\begin{equation}
  \langle O (x) O (0) \rangle_M\xrightarrow{x^0 \rightarrow i x^0 (1 \mp i
        \epsilon)}\langle O (x) O (0) \rangle_K  =  \frac{1}{[- (x^0)^2 - (x^1)^2 +
  (\vec{x})^2]^{\Delta}}\,.
\end{equation}

We can also consider direct analytic continuation from Euclidean space 
\begin{equation}
          \td s^2  =  (\td q)^2+ (q\td\psi)^2+(\td r)^2+(r \td\varphi)^2\,,\\
\end{equation}
to Klein space
\begin{equation}
          \td s^2  =  -(\td q)^2- (q\td\psi)^2+(\td r)^2+(r \td\varphi)^2\,,\\
\end{equation}
by complexifying $q$ and taking $q \rightarrow i q (1 \mp i \epsilon)$. Since $q>0$, we can see that there is only one possible branch point of correlators in Klein space that occurs at $x^2=0$, which is
\begin{equation}
  q =  r > 0\,.
\end{equation}
When $q < r$, starting from (\ref{eq:ConfEuc}), the correlator becomes
\begin{equation}\label{eq:ConfKlein0}
  \langle O (x) O (0) \rangle_E\xrightarrow{q \rightarrow i q (1 \mp i
        \epsilon)}\langle O (x) O (0) \rangle_K  =  \frac{1}{(- q^2 + r^2)^{\Delta}} \,.
\end{equation}
When $q > r$, the correlator has two branches as well
\begin{equation}
    \begin{aligned}
        \langle O (x) O (0) \rangle_E\xrightarrow{q \rightarrow i q (1 - i
        \epsilon)}\langle O (x) O (0) \rangle_K^{(1)}  = & \frac{1}{e^{i \pi \Delta} (q^2 -
        r^2)^{\Delta}}\,,\\
        \langle O (x) O (0) \rangle_E\xrightarrow{q \rightarrow i q (1 + i
        \epsilon)}\langle O (x) O (0) \rangle_K^{(2)}  = & \frac{1}{e^{- i \pi \Delta} (q^2 -
        r^2)^{\Delta}}\,.
    \end{aligned}
\end{equation}
In this work, we constructed the Klein correlators $\langle O (x) O (0) \rangle_K$ and $\langle O (x) O (0) \rangle_K^{(1)}$ as they are given by the Feynman path integral and interpreted as the $q$-ordered correlation functions in the canonical quantization formalism.

The causal structure of the spacetime is reflected in the correlation function,  as the branch points of the correlation functions are located at the light cone of the original point.  First, we see that the correlation function (\ref{eq:ConfEuc}) is definitely positive and only has a single pole at $x=0$, which means a ``light cone" in Euclidean space is the original point itself. In the Minkowski spacetime, consider the correlation function (\ref{eq:ConfMink0}) as a holomorphic function of complex $x^0$ with fixed $\vec x=(x^1,x^2,x^3)$, we have two branch points located at $x^0=\pm|\vec x|$ corresponding to the future and past light cones, respectively. In the Klein space, consider the correlation function (\ref{eq:ConfKlein0}) as a holomorphic function of complex $q$ with fixed $r$, we have only one meaningful branch point located at $q=r>0$ corresponding to the only light cone in Klein space. In short, the form of the correlation functions is determined by whether $x$ and the original point are space-like or time-like separated from each other: we have $\langle O (x) O (0) \rangle_E,\,\langle O (x) O (0) \rangle_M,\,\langle O (x) O (0) \rangle_K$ if they are space-like separated; we have $\langle O (x) O (0) \rangle_M^{(1)/(2)},\,\langle O (x) O (0) \rangle_K^{(1)/(2)}$ if they are time-like separated. The two different branches are traced back to the opposite $\epsilon$-prescription.

%\cb{(I can't see clearly where the causal structure appear. Please have more physical discussions.)}

\bibliographystyle{JHEP}
\bibliography{biblio.bib}
\end{document}